%% file: NRT_sample.tex
\documentclass[usenatbib]{mn2e}
\usepackage{cite}
\usepackage{epsfig}
\usepackage{color}
\usepackage{lscape}
\usepackage{natbib}
\usepackage{longtable}
\usepackage{journal_names}

\newcommand{\HI}{H\,{\sevensize{I}}} % MNRAS
\newcommand{\B}{$B$}
\newcommand{\R}{$R$}
\newcommand{\II}{$I$}
\newcommand{\J}{$J$}
\newcommand{\HH}{$H$}
\newcommand{\K}{$K_{\rm s}$}
\newcommand{\Ko}{$K^{\rm o}_{\rm s}$}
\newcommand{\ko}{$K^{\rm o}_{\rm s}$}

\newcommand{\koc}{$K^{\rm o,c}_{\rm s}$}
\newcommand{\Kd}{$K^{\rm o,d}_{\rm s}$}
\newcommand{\kd}{$K^{\rm o,d}_{\rm s}$}
\newcommand{\ad}{$a^{\rm d}$}
\newcommand{\ktw}{$K_{20}$}
\newcommand{\we}{$w1$}
\newcommand{\wz}{$w2$}
\newcommand{\wdr}{$w3$}

\newcommand{\ebv}{$E(B-V)$}
\newcommand{\ak}{A_{\rm K}}
\newcommand{\ab}{A_{\rm B}}
\newcommand{\hk}{$(H-K_{\rm s})$}
\newcommand{\jk}{$(J-K_{\rm s})$}

\newcommand{\hko}{$(H-K_{\rm s})^{\rm o}$}
\newcommand{\jko}{$(J-K_{\rm s})^{\rm o}$}

\newcommand{\hkoc}{$(H-K_{\rm s})^{\rm o,c}$}
\newcommand{\jkoc}{$(J-K_{\rm s})^{\rm o,c}$}

\newcommand{\nan}{Nan\c{c}ay}
\def\approxlt{\lower.2em\hbox{$\buildrel < \over \sim$}}
\def\approxgt{\lower.2em\hbox{$\buildrel > \over \sim$}}
\newcommand{\HII}{\mbox{H\,{\sc ii}}}

\def\corresp{\lower.1em\hbox{$\buildrel \wedge \over =$}}

\newcommand{\etal}{et~al.}

\newcommand{\eg}{{e.g.},\ }         % e.g. in italics
\newcommand{\ie}{{i.e.},\ }         % i.e. in italics
\definecolor{grey}{rgb}{0.5,0.6,0.7}

\title[{2M-ZoA galaxy catalogue description and analysis}]
      {A Zone of Avoidance catalogue of 2MASS bright galaxies. I. Sample description and analysis}

\author[A.C. Schr\"oder et al.]{A.C. Schr\"oder,$^{1}$\thanks{E-mail:anja@saao.ac.za}, 
W. van Driel$^{2,3}$, R.C. Kraan-Korteweg$^{4}$\\
$^{1}$South African Astronomical Observatory, PO Box 9, Observatory 7935, Cape Town, South Africa\\
$^{2}$GEPI, Observatoire de Paris, PSL Research University, CNRS, 5 place Jules Janssen, 92190 Meudon, France\\
$^{3}$Station de Radioastronomie de \nan, Observatoire de Paris, CNRS/INSU USR 704, Universit\'e d'Orl\'eans OSUC,\\ 
route de Souesmes, 18330 \nan, France\\
$^{4}$Department of Astronomy, University of Cape Town, Private Bag X3,\\ 
Rondebosch 7701, South Africa\\
}

\begin{document}

\date{Accepted....... ;}

\pagerange{\pageref{firstpage}--\pageref{lastpage}} \pubyear{2017}

\maketitle

\label{firstpage}

\begin{abstract}
We present a homogeneous 2MASS bright galaxy catalogue at low Galactic
latitudes ($|b| \le 10\fdg0$, called Zone of Avoidance) which is complete
to a Galactic extinction-corrected magnitude of \Ko\ $\le 11\fm25$. It also
includes galaxies in regions of high foreground extinctions (\ebv $>
0\fm95$) situated at higher latitudes. This catalogue forms the basis of
studies of large-scale structures, flow fields and extinction across the
ZoA and complements the ongoing 2MASS Redshift and Tully-Fisher surveys. It
comprises 3763 galaxies, 70\% of which have at least one radial velocity
measurement in the literature. The catalogue is complete up to star density
levels of $\log N_*/{\rm deg}^2 <4.5$ and at least for $\ak < 0\fm6$ and
likely as high as $\ak = 2\fm0$. Thus the ZoA in terms of bright NIR
galaxies covers only $2.5-4$\% of the whole sky. We use a
diameter-dependent extinction correction to compare our sample with an
unobscured, high-latitude sample. While the correction to the \K -band
magnitude is sufficient, the corrected diameters are too small
by about $4''$ on average. The omission of applying such a
diameter-dependent extinction correction may lead to a biased flow field
even at intermediate extinction values as found in the 2MRS survey. A
slight dependence of galaxy colour with stellar density indicates that
unsubtracted foreground stars make galaxies appear bluer. Furthermore,
far-infrared sources in the DIRBE/IRAS extinction maps that were not
removed at low latitudes affect the foreground extinction corrections of
three galaxies and may weakly affect a further estimated $\approxlt 20$\%
of our galaxies.
\end{abstract}


\section[]{Introduction} \label{intro} %1

Truly all-sky galaxy samples are difficult to acquire because of the
so-called Zone of Avoidance (hereafter ZoA). In the ZoA, Galactic dust
extinction as well as star crowding severely hamper not only the
identification of galaxies (where different wavelengths are affected by
different biases, \eg \citealt{KKLahav2000} and references therein) but
also subsequent follow-up observations.

While all-sky imaging surveys exist, like photographic plate surveys in the
optical such as the Palomar Observatory Sky Survey (POSS) and the
UK-Schmidt telescope (SERC) surveys (now available in digitised format at
DSS and SuperCOSMOS), and more recent CCD-based surveys in the NIR such as
2MASS (Two Micron All Sky Survey; \citealt{skrutskie06}), many other
surveys cover only one hemisphere (at least in one passband), \eg
DENIS (DEep Near-Infrared southern sky Survey; \citealt{Epchtein1997}), VHS
(VISTA Hemisphere Survey; \citealt{mcmahon13}), HIPASS (\HI\ Parkes All-Sky
Survey; \citealt{Meyer2004}), the DESI Legacy Imaging Surveys
  (\citealt{dey18}), and the upcoming LSST (Large Synoptic Survey
Telescope, \citealt{lsst}) survey. For extragalactic surveys and cosmic
flow field analyses, the objective is the {\it identification} of galaxies,
who can then be targeted for redshift measurements. Thus not only the
distinction between extended sources and point sources but also between
Galactic and extragalactic extended sources is tantamount. Automated galaxy
detection algorithms work well at high latitudes where the spatial density
of foreground stars is low and Galactic extended sources scarce, but they
are far inferior to the human eye in the crowded areas of the Galactic
Plane. Though such by-eye searches of the ZoA were done extensively using
photographic plates (\eg \citealt{KK2000a}, \citealt{WoudtKK2001} in the
south and \citealt{saito91} in the north), the modern CCD-based,
high-resolution surveys, like the UKIDSS GPS (Galactic Plane Survey;
\citealt{lucas08}), the Vista VVV (VISTA Variables in the V\'ia Lactea;
\citealt{minniti10}) and the DECam Plane Survey \citep{schlafly18},
have a depth and spatial resolution that renders such searches impossible
within a reasonable time span.

Furthermore, for a 3-dimensional (3D) view of the sky, spectroscopic
surveys are required which usually depend on a galaxy sample as input and
are thus, even when based on an all-sky imaging survey, usually restricted
to higher Galactic latitudes (\eg 2MRS: the 2MASS Redshift Survey,
\citealt{huchra12}; the 6dF Galaxy Survey, \citealt{jones09}; the Sloan
Digital Sky Survey SDSS, \citealt{york00}). Only so-called blind
\HI\ surveys with their great advantage of a 3D view of observed areas of
the sky cover the ZoA as well (\eg HIPASS, \citealt{Meyer2004}; HIJASS: the
\HI\ Jodrell All Sky Survey, \citealt{hijass}; EBHIS: the Effelsberg Bonn
\HI\ Survey, \citealt{kerp11}). Disadvantages of surveys in \HI\ are the
intrinsical weakness of the 21\,cm line emission, which leads to a
restricted coverage in redshift out to which galaxies can be detected, as
well as the paucity of \HI\ in certain types of galaxies, in particular of
elliptical galaxies and early-type spirals that often define the centres of
massive clusters. The advantages are that galaxies can be detected
independent of foreground extinction and star crowding (though their
detection is affected by the presence of radio continuum sources which are
more frequent in the Galactic Plane, \citealt{staveley16}). Thus, blind
\HI\ surveys of the ZoA have been the tool of choice in mapping large-scale
structures in this area (\eg HIZOA: the \HI\ Zone of Avoidance survey,
\citealt{staveley16}; HIZOA-NE: the HIZOA Northern Extension survey,
\citealt{Donley2005}; ALFAZOA: the Arecibo L-band Feed Array Zone of
Avoidance Survey, \citealt{alfazoa}; EZOA: the EBHIS Zone of Avoidance
survey, Schr\"oder \etal , in prep.).

Truly all-sky galaxy samples that are homogeneous are therefore difficult
to compile but they are important to arrive at a complete and unbiased 3D
map of the local universe required to explain the dynamics in the local
universe (\eg \citealt{qin18} using 2MTF and 6dFGSv, \citealt{courtois17}
using CosmicFlows3, \citealt{lavaux16} using 2M++, \citealt{erdogdu06b}
using 2MRS) including structures such as the Great Attractor
\citep{Dressler1987}, the Perseus Pisces Complex \citep{giov82} or the
Local Void \citep{Tully2008}.  Even though the gap due to the ZoA is
considerably smaller in the NIR than in the optical, the incompleteness in
the 3D galaxy mass distribution due to the ZoA still leads to new
discoveries (\eg the Vela Supercluster, \citealt{KK2017},
\citealt{sorce17}; a rich cluster in the Perseus-Pisces filament,
\citealt{ramatsoku16}) and still has an impact on the studies of the cosmic
flow fields where controversies remain in modelling the dipole of the
cosmic microwave background (\eg \citealt{KKLahav2000},
\citealt{Loeb2008}). For more information, see our Paper II \citep{kk18}.

Another advantage of defining a homogeneous galaxy sample that includes the
high extinction areas of the ZoA is that it allows the study of systematic
effects which are caused by an incomplete understanding of the impact of
extinction on galaxy sample selections that are based on magnitudes or
size. Even at higher Galactic latitudes ($|b|>10\degr$) there are areas of
higher extinction on the sky (\eg the Orion Nebula or the Taurus Molecular
Cloud) where incompletely corrected galaxy magnitudes could lead to
systematic biases in an all-sky analysis (see our discussion on Galactic
extinction corrections in Sec.~\ref{selection}).

We therefore compiled a magnitude-limited sample of 2MASS galaxies in the
ZoA to contribute to a homogeneous `whole-sky' 2MRS for large-scale
structure studies in the nearby Universe, as well as to the `whole-sky'
2MASS Tully Fisher Relation Survey (2MTF; \eg \citealt{masters14}) of
inclined spiral galaxies to study cosmic flow fields. Although both these
major surveys intend to map the full three-dimensional distribution of
  galaxies in the nearby universe they have excluded the inner part of the
ZoA ($|b| < 5\degr$), and are not fully complete for its outer part ($5 <
|b| < 10\degr$). Both are based on the 2MASS extended source catalogue
(2MASX; \citealt{Jarrett2004}) and have a \K-band magnitude completeness
limit of $11\fm75$ and $11\fm25$, respectively. Though the 2MASX catalogue
itself {\it includes} the ZoA, its reliability drops off quickly at low
latitudes ($|b| < 10\degr$; \citealt{Jarrett+2000b}, \citealt{huchra12}).

Contrary to the 2MRS magnitude limit of \Ko\ $< 11\fm75$, we adopted a
brighter limit of \Ko\ $< 11\fm25$, which is the same as the limit of the
first 2MRS data release \citep{huchra05} as well as that of the 2MTF
survey. At this extinction-corrected magnitude limit the 2MASX completeness
level remains fairly constant across the ZoA \citep{huchra05} apart from
the Galactic Bulge. The completeness would drop substantially towards lower
latitudes, however, if the extinction-corrected magnitude limit is lowered
to include fainter ZoA objects, due to the steep increase in the number of
faint foreground stars and the related rise in sky background close to the
Galactic Plane \citep{Jarrett+2000b}. Furthermore, deep in the ZoA the \K
-band extinction corrections can reach values of up to 1 magnitude, which
implies that the apparent uncorrected magnitudes are quite faint which will
affect both the accuracy of their photometry as well as their morphological
classification.  The \Ko\ $= 11\fm25$ limit corresponds to a $B$-band limit
of $14\fm75$ for an average spiral galaxy colour of $B-K = 3\fm5$
\citep{jarrett03, Jarrett2000}, and is thus comparable to the completeness
limit of most optically selected nearby galaxies whole-sky surveys
\citep{KKLahav2000}.

Originally, our intention was to observe 2MASS galaxies in the {\it
  northern} ZoA that had no prior redshifts in the 21\,cm line, using the
100\,m class \nan\ Radio Telescope (NRT). So far no systematic \HI\ surveys
of galaxies have been conducted over the northern ZoA, unlike in the south
where pointed as well as blind \HI\ surveys were made using the Parkes
radio telescope (\eg \citealt{kraan02}; \citealt{donley05};
\citealt{schroeder09}; \citealt{staveley16}) or at medium declinations
using the Arecibo telescope (\eg \citealt{pantoja94}, \citealt{pantoja97}).
And with an interstellar extinction in the \K-band ($\lambda\, 2.2\,\mu$m)
11 times smaller than in the $B$-band, absorption remains relatively modest
for galaxies in this passband over most of the northern ZoA, despite many
being invisible on optical Palomar Sky Survey (DSS, Digital Sky Survey)
images, while confusion due to star-crowding in the NIR is minimal for most
of the northern ZoA. We thus started with a list of about 1100 2MASX
objects compiled by J.\ Huchra (priv.\ comm.)  mostly at $|b|<5\degr$ and
$\delta > -40\degr$ (so as to be observable with the NRT), then formalised
the selection criteria to include all objects from the 2MASS NIR extended
objects catalogue \citep{Jarrett+2000a} of the 2MASS survey at
$|b|<10\degr$. For completeness reasons and to finally have a homogeneous
all-ZoA (and thus all-sky) galaxy sample, we extended our sample to cover
the southern ZoA as well as high-latitude high extinction areas to fully
complement the 2MRS.

This paper is the first of a series. A pilot project based on 197 2MASX ZoA
galaxies was presented by \citet{vandriel09}. In Paper II \citep{kk18} --
(Kraan-Korteweg et al. 2018) we present the results of our NRT \HI\ survey
of a thousand 2MASX ZoA galaxies, and Paper III (Schr\"oder et al., in
prep.) will address the Galactic foreground extinction in the ZoA. Further
publications will include observations at more southern declinations using
the Parkes radio telescope, \HI\ observations of galaxies {\it with}
redshifts (mainly in the optical, but also of those galaxies with lower
grade \HI\ spectra) that are eligible for the TF analysis, as well as the
flow-field analysis using the NIR Tully-Fisher relation.

This paper is structured as follows: in Sec.~\ref{sample}, our sample
selection and extraction is described and in Sec.~\ref{selection} details
and limitations of the selection criteria are discussed.
Sec.~\ref{results} presents the catalogue, and its properties are discussed
in Sec.~\ref{prop}. Section~\ref{summary} gives the summary and
discussion. A supplementary catalogue of ZoA galaxies which need to be
included in a sample when optimised extinction corrections are applied is
presented in the appendix.

\section{The sample} \label{sample} %2

As mentioned in the introduction, our foremost aim is to complement the
2MRS and 2MTF efforts by obtaining redshifts of bright galaxies in the ZoA, that
is, in the latitude range not covered by the 2MRS. The 2MRS sample
selection criteria are:
\begin{enumerate}
\item \Ko\,$ \le 11\fm75$;
\item \ebv\,$ < 1\fm0$;
\item also detected at \HH-band;
\item $|b| > 5\fdg0$ for $30\degr < l < 330\degr$; $|b| > 8\fdg0$
  otherwise.
\end{enumerate}
We selected our sample using \K-band magnitudes corrected for Galactic
extinction, \Ko. To be as close as possible to the 2MRS magnitude-selection
criterion, we extracted the \ebv\ Galactic extinction values directly from
the DIRBE/IRAS maps by \citet[][hereafter SFD98]{schlegel98} and did {\it
  not} correct for the factor of 0.86 applied by \citet[][hereafter
  SF11]{schlafly11}.

We defined two samples of galaxy candidates from the 2MASX catalogue: the
main sample, which we will refer to as the ZOA sample, and the
complementary EBV sample at higher latitudes. Their selection criteria are
as follows:
\begin{description}

\item{\it Both samples}: unlike the 2MRS, we did not require detection in
  the \HH-band but chose a slightly brighter magnitude limit than the 2MRS,
  \ie \Ko$\le 11\fm25$.

\item{\it ZOA sample}: objects at $|b| \le 10\fdg0$ at all Galactic
  longitudes;

\item{\it EBV sample}: all objects at $|b| > 10\fdg0$ with \ebv $ >
  0\fm95$. The overlap between our \ebv\ limit and that of the 2MRS (\ie
  $1\fm0$) covers any small variations in the determination of \ebv;

\item{\it Galaxy candidates:} both samples (particularly at high
  extinctions and high star-density regions) are contaminated with Galactic
  Nebulae and blended stars. We therefore visually inspected all 2MASX
  objects to decide whether they are likely galaxies or not (for further
  details, see Sec.~\ref{extr}).

\end{description}

To search for radial velocity information, we cross-correlated both our
samples with the 2MRS catalogue as well as with the NASA/IPAC Extragalactic
Database (NED)\footnote{http://ned.ipac.caltech.edu/} and the HyperLeda
database\footnote{http://leda.univ-lyon1.fr/}.

\subsection{Sample extraction } \label{extr} %2.1

\begin{table}  % 1
\centering
\begin{minipage}{140mm}
\caption{{Extinction correction factors in the NIR } \label{tabext}}
\begin{tabular}{clll}
\hline
Passband & \multicolumn{1}{c}{$\lambda$} & $A_\lambda$ / $\ab$ & Reference \\
         & ($\mu$m)  &                        &            \\
\hline
\J    & $\phantom{1}1.25$  & 0.21   &  \citealt{fitz99}  \\ 
\HH   & $\phantom{1}1.65$  & 0.13   &  \citealt{fitz99}  \\ 
\K    & $\phantom{1}2.15$  & 0.09   &  \citealt{fitz99}  \\ 
\we   & $\phantom{1}3.4$   & 0.057  &  \citealt{schlafly16}  \\ 
\wz   & $\phantom{1}4.6$   & 0.045  &  \citealt{schlafly16}  \\ 
\wdr  & $12.0$             & 0.086   &  \citealt{davenport14} \\ 
\hline
\end{tabular}
\end{minipage}
\end{table}

Our catalogue is based on the 2MASX catalogue\footnote{See the 2MASS
  All-Sky Extended Source Catalog (XSC) as found online at
  http://irsa.ipac.caltech.edu/cgi-bin/Gator/nph-dd}. For the ZOA sample,
we first downloaded all sources with $|b| \le 10\fdg0$, using the default
SQL constraints given in the webform, \ie deselecting sources with
contamination and confusion flags, which are usually marked as
  ``junk'' ({\tt cc\_flg}=~`$a$' for known artefacts and~`$z$'
for small detections in close proximity to large galaxies). This
resulted in 146\,174 sources. We extracted extinction values for each
source using the SFD98 maps supplied through their
webpage\footnote{http://w.astro.berkeley.edu/$\sim$marc/dust/} and the {\tt
  dust\_getval} script,
setting the {\tt interp} parameter to `y'. We determined the extinction
correction in the \K-band using $R_{\rm B} = \ab $/\ebv\ = 4.14 and the
conversion $\ak = 0.09 \ab $ (\citealt{fitz99}\footnote{we prefer
    \citet{fitz99} over \citet{Cardelli+1989} since it has been found to
    represent the \HH -band extinction better, as discussed in Paper III}). Table~\ref{tabext} lists the
extinction correction factors compared to the \B -band value for all
passbands used in this publication. We applied this \K-band extinction
correction to the 20 mag arcsec$^{-2}$ isophotal magnitude, \ktw , and cut
the list at the corrected \Ko\ $\le 11\fm25$, which resulted in 6913
sources.

For the higher-latitude EBV sample, we first extracted all 2MASX-sources at
$|b|>10\fdg0$, obtained their \ebv\ values and applied a cut at \ebv $ \ge
0\fm95$, leading to 4248 sources. Applying the \Ko\ $\le 11\fm25$ limit
results in 502 sources.

The 2MASX sources in the ZoA are not all extragalactic in nature but also
comprise various kinds of Galactic sources, including multiple stars and
small stars on diffraction spikes \citep{Jarrett+2000b}. Although the 2MASX
catalogue provides a visual verification score `{\it vc}' which separates
truly extended sources from stars and artefacts, it does not separate out
candidate galaxies from Galactic sources, and we visually inspected all
7415 sources. Even to the human eye, galaxy classification in the ZoA is
not straight forward. A major problem are the ambient high stellar
densities: Superimposed stars hinder the identification of a halo-like disc
structure around the core region, while the ubiquitous unresolved stars in
the field increase the surface brightness of the background which
diminishes the visibility of the low surface brightness parts of the
galaxian discs.

We therefore decided to use as much information as possible. We downloaded
multi-wavelength images from several online archives; the particular
advantages of using specific data sets are explained below. The image sets
are, in order of increasing wavelength:

\begin{itemize}

\item \B -band images from the SuperCOSMOS Sky
  Surveys\footnote{http://www-wfau.roe.ac.uk/sss/};

\item \R - and \II -band images from the Digitized Sky Survey (DSS,
  2$^{nd}$
  generation)\footnote{http://www3.cadc-ccda.hia-iha.nrc-cnrc.gc.ca/en/dss/
  };

\item 2MASS \J - and \K -band Atlas
  images\footnote{http://irsa.ipac.caltech.edu/applications/2MASS/IM/batch.html}
    as well as the $JHK$
    composite\footnote{http://irsa.ipac.caltech.edu/applications/2MASS/PubGalPS/}
    images;

\item UKIDSS\footnote{http://surveys.roe.ac.uk/wsa/} \K -band images (where
  available);

\item VISTA Imaging Public Surveys\footnote{http://horus.roe.ac.uk/vsa/} \K
  -band images (where available);

\item WISE\footnote{http://irsa.ipac.caltech.edu/applications/wise/} NIR
  images at $3.4-22\mu$m wavelength;

\item DIRBE/IRAS dust maps (SFD98) at the location of the 2MASX source, at
  $100\mu$m wavelength.

\end{itemize}

In the selection process we furthermore used the extinction information
itself as well as the extinction-corrected NIR-colours.

The combination of all this information was most helpful in classifying the
sources and compiling a reliable galaxy sample. The series of images from
the optical to the NIR ($\lambda$$0.4-2.2\mu$m) allowed to see the source
`in sequence' of decreasing foreground extinction levels, while the
absolute extinction values themselves were useful to estimate the {\it
  expected} change in appearance over this wavelength range. The \B - and
\R -band images were best for identifying Galactic dark clouds in the area
around a source, which may indicate the presence of Young Stellar Objects;
these also appear bright in the \K- band. Dark clouds may also imply
small-scale variations in the foreground extinction. UKIDSS and VISTA \K
-band images were of great help due to their higher sensitivity and spatial
resolution, but not all areas of the sample regions are covered by these
surveys. Although the shape and colour of objects as seen in WISE images,
which have lower resolution than the previously mentioned surveys, can help
to distinguish between Galactic and extragalactic objects, it appears
better to separate, \eg Planetary Nebulae from galaxies using 2MASX colours
(see Sec.~\ref{colours}). The DIRBE/IRAS dust maps helped to identify
infrared point sources which were not removed from the maps at low Galactic
latitudes ($|b|<5\degr$) and which skew the extinction values around that
location. Finally, the galaxy colours form a tight distribution in NIR
colour-colour plots \citep{Jarrett2000}. Due to Galactic extinction, the
colours become redder and move along a well-defined reddening path with the
width of the intrinsic colour dispersion of galaxies. Hence, even in cases
where the extinction is not measured correctly due to, \eg spatial
variation being smaller than the resolution of the SFD98 dust maps (6
arcminutes) or a near-by FIR source not being removed from the maps, a
galaxy will lie on this reddening path. Any object with colours outside
this path is therefore unlikely to be a galaxy (but see also
Sec.~\ref{prop} on the discussion of the properties of the catalogue).

\subsection{Radial velocities} \label{vel} % 2.2

We searched the 2MRS catalogue and the NED and HyperLeda database for
publicly available redshifts. We found that only 34\% of the galaxies in
both our ZOA and EBV samples are not in the 2MRS catalogue. Furthermore, in
preparation of our \HI\ observing campaigns we distinguished between
optical and \HI\ measurements. The redshift information was also compared
with the galaxy candidate classification which was adjusted where
appropriate (\eg a high velocity measurement could confirm an object to be
a quasar). We note, however, that not all redshift information was found to
be reliable. In optical spectroscopy, lines in the spectrum as well as the
target itself (in particular in high obscuration areas) can be
mis-identified -- where we identified such a problem, the 2MRS catalogues
were updated accordingly (Macri 2016, priv. comm.). In case of
\HI\ velocities the relatively large beam-width of radio telescopes
($3\farcm5-23\arcmin$ for the single-dish instruments used) made it
possible that another galaxy was detected in the beam instead of, or
together with, the target galaxy (see, for example, the notes in the
appendix of Paper II). We mark such cases with a question mark
(questionable ID) or colon (questionable value) in our catalogue.

Due to our observations as well as ongoing activities by the 2MRS team and
other colleagues, we also set flags for galaxies with recently determined
redshifts that are as yet unpublished. These are (a) 2MRS optical data
(Macri, priv.comm.), (b) HIZOA Galactic Bulge survey Parkes detections
\citep{kk18}, (c) ALFAZOA survey Arecibo detections (see
\citealt{mcintyre15} for first results), (d) EBHIS-ZoA survey Effelsberg
detections (Schr\"oder \etal, in prep.), (e) the NRT detections from Paper
II, and (f) our Parkes (PKS) detections (Said \etal, in prep.).

\subsection{Galaxy types} \label{type} % 2.3

To prioritise target galaxies for our \HI\ observing campaigns, we decided
to classify them according to morphological type. This is more difficult
than in the optical because the disc of a spiral galaxy appears much
smoother in the NIR \citep{Jarrett2000}. Moreover, the high foreground
extinction in the ZoA affects lower surface brightness discs more than
bulges, and the more sophisticated classification methods presented in
\citet{Jarrett2000} are less reliable in the ZoA. We therefore introduced a
simplified morphological classification and distinguished purely between
the presence of a disc and its absence. We defined four classes in order of
decreasing likelihood of a disc being present: D1 = obvious disc, D2 =
highly probable disc, D3 = possible disc and D4 = no noticeable disc. We
set a no-flag `n' when it was not at all possible to estimate the type, \ie
where one or more bright stars were too close to discern a disc, where the
object was too small, or where the extinction was so high that all of the
disc might be obscured. Note that we used all available images for this
classification; the \B - and \R -band images were especially helpful when
the extinction was low.

\subsection{The ZOA sample} \label{zsample} % 2.4

Based on the above-mentioned sample cleaning process, we gave each object a
flag from 1 to 9, indicating its likelihood of being a galaxy, with 1 being
the most likely. Table~\ref{tabflags} summarises the flags and gives the
numbers of objects found. Flags 1\,--\,4 indicate galaxies, whereas flags
6\,--\,9 indicate non-galaxies; the four objects with flag 5 (`unknown')
are kept in the galaxy sample as galaxy candidates until more information
becomes available. Thus of the 6913 objects in our ZOA sample, 3675 are
classified as galaxies.

In 32 cases a 2MASX object was labelled by us as "not a galaxy" but is in fact a
detection of a galaxy near the edge of an Atlas image where its centre was
obviously not determined correctly. In all cases the galaxy was also
detected on the adjacent (overlapping) image with correctly centred
coordinates. We retained the latter in our galaxy sample, while the former,
offset-detected galaxy, received the object flag~9 in combination with an
offset flag~`e' (for `edge').

\begin{table}
\centering
\begin{minipage}{140mm}
\caption{{Flags for identifying 2MASX sources as galaxies} \label{tabflags}}
\begin{tabular}{llrr}
\hline
Flag & Galaxy class & ZOA sample & EBV sample \\
\hline
1+2 & definitely         & 3609 &  84 \\
3 & probably             &   42 &   1 \\
4 & possibly             &   20 &   3 \\
5 & unknown              &    4 &   0 \\
6 & low likelihood       &   29 &   7 \\
7 & unlikely             &  340 &  87 \\
8+9 & no                 & 2869 & 320 \\
\hline
total &                  & 6913 & 502 \\
\hline
\end{tabular}
\end{minipage}
\end{table}

We compared our results with the 2MRS sample for $|b| \le 10\degr$, using
the version from 16 December2011\footnote{http://tdc-www.harvard.edu/2mrs/} 
\citep{huchra12}. There are in fact two 2MRS catalogues listed, the `main'
catalogue, defined according to the aforementioned sample criteria (see
Sec.~\ref{extr}), and the `extra' catalogue which contains entries outside
those criteria (and mainly at lower Galactic latitudes) but which is not
complete. Each catalogue is further divided into sub-catalogues of objects
which have no or wrong velocity measurements, have bad photometry, or were
`rejected' as they are believed not to be galaxies.

The agreement in galaxy classification between the 2MRS catalogues and our
ZOA sample is excellent ($>99$\%): of the 6913 objects in our ZOA sample
(which include 3675 galaxies), 1778 are in the main 2MRS catalogue (all of
which we also classified as galaxies) and 20 were rejected by 2MRS (two of
which we labelled as galaxies while a further 12 are `edge' detections),
whereas the 2MRS `extra' catalogue lists 613 of our objects (610 of which
we classified as galaxies) and rejected a further 14 of our objects (of
which 9 are labelled as galaxies).
This means that in our ZOA galaxy catalogue we retained 11 2MRS rejects as
galaxies but rejected three of their galaxies as non galaxian. For 20 of
our objects a low velocity confirmed their Galactic nature. In summary, we
have 1276 galaxies in our ZOA sample that are not in the 2MRS catalogues.

\subsection{The EBV sample} \label{esample} %2.5

The EBV sample comprises 502 objects of which 88 are galaxies (see
Table~\ref{tabflags}). The main 2MRS catalogue lists 16 of the objects as
galaxies (15 of which we also classified as galaxies) and 14 as rejected
(which we also all rejected), whereas the 2MRS `extra' catalogue lists 57
of our objects, all of which we also classified as galaxies, and none as
rejected. This means that we rejected only one of the 2MRS galaxies as
non-galaxian in our EBV catalogue, while only 16 galaxies from our EBV
sample are not in the 2MRS catalogues.

\section{On selection criteria in the ZoA} \label{selection} %3

Selection criteria that are based on measured parameters are subject to the
uncertainties in those parameters. Since ZoA galaxies on average have
higher measurement errors (\eg due to star crowding) and are also subjected
to various corrections, we will discuss in the following individual error
sources and their effects on the sample selection. This will be
particularly important when combining any kind of ZoA catalogue with a
high-Galactic latitude catalogue.

\subsection{Quality of the photometry} \label{badphot} % 3.1

The sample selection obviously depends on the quality of the 2MASX \K -band
magnitude. We noted different kinds of issues with photometric data: \\

(1) We list the following three photometry flags as used in the 2MRS
catalogues, for a total of 38 of our galaxies: {\it rep/add} (flag `a' in
our catalogue) indicating that improved photometry is given in the 2MRS
{\it add}-catalogue; {\it flr} (our flag `p') where the photometry is
deemed equally compromised but the galaxy has not been reprocessed yet;
{\it flg} (our flag `f') for objects with poor photometry which could
benefit from reprocessing. \\

(2) We give the improved-photometry flag from John Huchra's original list
(priv.\ comm.), for 22 objects (8 of which are galaxies): they are flagged
as `c' (for corrected) in our catalogue. For consistency throughout, we do
not use the improved photometry for the sample definition. Eight of these
objects (two of them galaxies) would fall below our magnitude cut-off with
the improved photometry. \\

(3) We give an offset flag for sources that are centred on a superimposed
star rather than the galaxy bulge: while visually inspecting all the
images, we flagged all such sources with an `o' (for offset) in our
catalogue. We assume that their photometry (including those based on
central apertures) may be compromised. Out of the 263 cases, 166 are
galaxies.  Sixteen of these galaxies are also flagged for bad photometry in
the 2MRS catalogues (12 of which as severe cases; see Item 1.) while one is
listed as rejected. \\

(4) Other 2MASX photometry issues: since the 2MASX photometry pipeline is
fully automated\footnote{except for the Large Galaxy Atlas}, we
expect that problems will occur in the crowded areas of the ZoA. For a
quick general check we selected a random sample of a few hundred galaxies,
mainly at lower latitudes, and extracted their 2MASX parameters r\_k20fe
(radius), sup\_ba (axial ratio) and sup\_phi (position angle) which
describe the fiducial elliptical aperture from which the isophotal
magnitude was determined. We then compared these visually with the actual
images. We find that though the aperture mostly agrees well with the visual
size and axial ratio of the galaxy (except in areas of high stellar
densities, \ie near the Galactic bulge), the position angle was often
affected by nearby (unsubtracted) stars, background variation etc., and in
some cases was rotated by a full 90 degrees. \\

The above-mentioned cases have diverse effects on the sample
definition. The worst effect seems to occur in the case of centring on a
superimposed star: this means the star was not subtracted and thus may
dominate the photometry (while, in some cases, the bulge of the galaxy was
treated instead like a point source to be subtracted). As a consequence,
the galaxy may in fact be fainter than the limiting magnitude. In addition,
the colours of the source may reflect those of the star and not of the
galaxy itself. However, only a small percentage (\ie 4\%) of our galaxy
sample is affected by this problem. Similarly, only 1.2\% of the sample is
flagged for bad photometry (Items 1. and 2. in the list). Only a few of
these would fall below the limiting magnitude of our sample with improved
photometry.

\begin{figure} 
\centering
\includegraphics[width=0.45\textwidth]{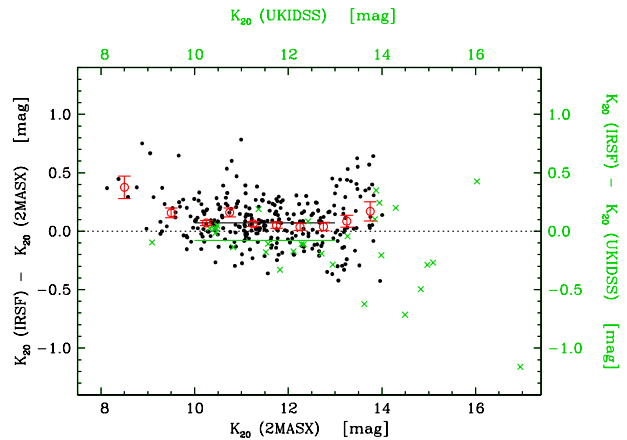}
\caption{Difference between \ktw\ magnitudes from IRSF and 2MASX (black
  dots) and IRSF and UKIDSS (green crosses) as a function of 2MASX
  \ktw. The average difference over the $10^{\rm m} - 13^{\rm m}$ range in
  \ktw\ is indicated by a horizontal black and green line, for
  IRSF\,--\,2MASX and IRSF\,--\,UKIDSS, respectively. The red open circles
  indicate binned means and errors of IRSF\,--\,2MASX magnitudes between 
  $8^{\rm m}$ and $14^{\rm m}$ in
  \ktw . For comparison, the dotted line indicates a
  zero difference.
}
%\vspace{2cm}
\label{compirsfplot}
\end{figure}

The effects of the cases where the automated photometry failed (Item 4.)
can go either way: where the position angle is wrong or the size too small
the \K -band magnitude will be (slightly) underestimated, while in cases
where the size was overestimated (due to non-subtracted star(s) near the
galaxy) the \K\ magnitude will be overestimated. In addition, other
non-subtracted stars within the aperture (but not affecting the position
angle), will also lead to \K -band magnitudes that are too bright.

An overall comparison of the 2MASX photometry with that from deeper NIR
images obtained at the InfraRed Survey Facility (IRSF) telescope at the
South African Astronomical Observatory was made by \citet{said16b}: see
their Figure~7 for a comparison of the 2MASX isophotal \ktw\ and IRSF
magnitudes, and also their Figure~8 for a comparison between IRSF and
UKIDSS magnitudes. They report good agreement for the range $9^{\rm m} -
14^{\rm m}$. Using the same datasets, we find (see Fig.~\ref{compirsfplot})
the 2MASX \ktw\ magnitudes to be on average $0\fm071\pm0\fm012$ brighter
than those of the deeper IRSF images (black dots) in the \ktw\ range
$10^{\rm m} - 13^{\rm m}$ ($N=207$). This value compares well with the
value of $0\fm08\pm0\fm01$ value obtained by \citet{williams14} for 102
galaxies in the precursor project. Furthermore, we find that IRSF
\ktw\ magnitudes are brighter by $0\fm078\pm0\fm035$ compared to the UKIDSS
measurements (green crosses). This results in a total offset of
$0\fm15\pm0\fm04$ between 2MASX and UKIDSS photometry.

As argued by \citet{williams14}, the obvious explanation is that unresolved
faint stars are the cause of this offset. Nevertheless, we do not find any
dependence of the offset with stellar densities (nor with colour or
extinction) for this data set. We thus conclude that the effect of
unresolved stars on the photometry is already present at the moderate
stellar density level of $\log N_*/{\rm deg}^2$ = 3.5 (\ie the lower limit
for this data set). Further investigation is required to determine at which
stellar density this effect becomes important.

In addition, \citet{andreon02} and \citet{Kirby+2008} report that 2MASX
isophotal fluxes in general are {\it under}estimated by more than 20\%,
which is assumed to be due to the short exposure times used by 2MASS (this
also affects any luminosity functions derived from 2MASX isophotal fluxes,
as discussed in detail by \citealt{andreon02}). This would mean that 2MASX
galaxies in the ZoA are too bright by at least $0\fm35$ magnitudes as
compared to higher latitude 2MASX galaxies (assuming the UKIDSS GPS
photometry, used as reference, is accurate, and because the UKIDSS
photometry are calibrated using 2MASS stars) since a 2MASX galaxy is
$0\fm20$ too faint compared to deeper photometry but, in the ZoA, they are
found to be $0\fm15$ too bright instead. This needs to be taken into
account when combining our sample with the 2MRS and 2MTF samples.

Finally, we can estimate the uncertainty in the limiting magnitude of our
sample to be $0\fm15$ based on the scatter of the IRSF\,--\,2MASX magnitude
difference in the 2MASX magnitude bin $11^{\rm m} - 11\fm5$.

\subsection{Correction for Galactic extinction } \label{badext} % 3.1

We used the DIRBE/IRAS FIR maps (SFD98) to extract extinction information;
these maps, however, are uncalibrated at low Galactic latitudes
($|b|<5\degr$). Among others, \citet{schroeder07} used NIR galaxy
photometry of a cluster around PKS\,1343--601 to derive a correction factor
of 0.87 to \ebv\ at low latitudes, and SF11 determined that the whole-sky
SFD98 maps had to be recalibrated by applying a factor of 0.86. We prefer
SFD98 over other maps in the literature (\eg \citealt{planck16}) since they
are used widely, and our values are thus consistent with 2MRS and 2MTF
surveys.  In a forthcoming paper (Schr\"oder \etal , Paper III) we will
present a detailed comparison.

We decided {\it not} to use the SF11 correction factor of 0.86 in our
sample {\it selection} so as to remain consistent with the 2MRS sample
definition. This correction factor can always be applied at a later stage
and the sample will still be complete. Note that only 204 galaxies ($<6$\%)
would become fainter than the magnitude cut-off if this correction factor
were applied.

Around some objects the foreground extinction shows a strong, local spatial
variation which is not reflected in the DIRBE/IRAS maps given their
relatively coarse pixel size of $6\arcmin$$\times$$6\arcmin$. Wherever the
$10\arcmin$$\times$$10\arcmin$ \B - or \R -band images showed such obvious
small-scale variations, we set an extinction flag. Sometimes a variation
across the image was discernible but without the sharp edges typical of a
dark cloud; in these cases we set a flag indicating that the extinction
{\it could} be wrong. Note that this was done systematically for galaxies
only. A total of 217 galaxies are affected; these flags are entirely based
on visual impression and are not complete. Notable is also that the
extinction is more likely to fluctuate on small scales where it is
intrinsically high (that is, in the area of dark clouds or Galactic
nebulae).

In regions where extinction shows strong spatial variations, the true
extinction may be either higher {\it or} lower, and the effect of
extinction on the sample selection cannot be estimated. Based on the
above-mentioned numbers, though, the effect is deemed to be small.

\subsection{Additional Galactic extinction corrections} \label{diamcorr}

For determining isophotal magnitudes the applied Galactic foreground
extinction correction is not sufficient (\citealt{Cameron1990},
\citealt{riad2010}): the dust in the Milky Way forms a screen which lowers
the surface brightness of any galaxy behind it, and therefore reduces their
observed radius (\eg at the 20$^{\rm th}$ mag arcsec$^{-2}$ level). We
therefore need to correct the observed \ktw\ isophotal diameter for this
effect and hence also the isophotal magnitude.

We have followed the method outlined by \citet{riad2010} using the 2MASX
central surface brightness (k\_peak), since 2MASX does not supply a disc central
surface brightness as required for an optimal correction. The method also
requires knowledge of the morphological type, since the surface brightness
profiles of elliptical and spiral galaxies (or bulges and discs)
follow different laws. However, morphological classification is more
difficult in the NIR compared to the optical, and even more so in the ZoA
where the disc is partly or even fully obscured. We therefore decided to
apply the correction for spiral galaxies to all objects in our catalogue, 
keeping in mind that (a) only about 20\% of all galaxies are ellipticals,
(b) one could use the disc class parameter (Sec.~\ref{type}) to distinguish
between spirals and ellipticals (see discussion below), and (c) we are
mainly interested in the Tully--Fisher relation which only applies to
spiral galaxies. To optimise this correction, though, we also applied the
correction factor 0.86 to the \ebv\ values (SF11).

The effect of the additional extinction correction is complex: the SF11
correction to the \ebv\ values makes galaxy magnitudes fainter, but the
diameter correction is additive and thus makes magnitudes
brighter. Therefore, some previously too faint objects would now lie above our
$11\fm25$ magnitude limit, and some sample objects would become fainter. To
investigate this effect we flagged all objects from our sample that would
be {\it excluded} when using the fully corrected magnitudes
\Kd\ (Col.~7d), and we extracted all objects from the 2MASX catalogue that
would be {\it added} to the sample (presented in the appendix). We thus
would have an additional 70 and 1 galaxies in the ZOA and EBV samples,
respectively, but would also lose 29 and 0, respectively, which means that
the sample size would increase marginally by about 1\% due to an improved
extinction correction.

We corrected the major-axis diameters $a$ and the \K -band magnitudes for
the diameter-dependent extinction, but not the colours: this because the
2MASX \J - and \HH -band `isophotal' magnitudes, from which the colours are
derived, refer to the \K -band isophotal fiducial elliptical aperture and
not a {\it fixed} isophote in the \J - and \HH -bands. Since the NIR
colours of galaxies change only slowly with radius (especially in the outer
parts; \citealt{jarrett03}), the conventional extinction corrected colours
out to the uncorrected isophotal radius will be sufficient for our purposes
and, in particular, they do not affect the sample definition. To be
consistent with the \Kd\ magnitudes though, we did apply the correction
factor 0.86 (SF11) to the colours (denoted \hkoc\ and \jkoc\ ).

For elliptical galaxies the corrections are smaller than for spirals due to
their steeper surface brightness profiles. Since we applied the correction
for spirals to all objects, the \Kd -sample size would decrease slightly if
we could properly distinguish elliptical from spiral galaxies. We preferred
not to make such a distinction using the disc parameter (see
Sec.~\ref{disctype} for more details) for the following reasons: (a) The
radial profiles of spiral galaxies (and thus the correction) also vary with
morphological subtype, hence the uncertainty in the correction at higher
extinctions quickly becomes significant. (b) The uncertainty in the
correction method we used is larger than for the optimal correction method
which uses the disc central surface brightness (see the detailed discussion
in \citealt{riad2010}). (c) The disc parameter is only reliable in those
cases where a disc is noticeable; it is more uncertain when discriminating
elliptical from early-type spiral galaxies (classes `D3' and `D4'), in
particular at higher extinctions where only bulges of spiral galaxies might
be visible (and thus a correction is technically not applicable
anyway). (d) It is not clear what kind of correction to use for the disc
classes `D3' and `n' which make up about 18\% of the galaxy sample. In
other words, using the disc parameter for the extinction correction would
only introduce an unknown uncertainty which may or may not be larger than
the correction applied.

The caveats outlined here indicate that we need to be cautious when
applying this additional correction. We therefore restricted ourselves to
the conventional extinction correction. The additional galaxies presented
in the appendix will be used for further discussions on selection effects
in later analyses; they will be important also for the TF analysis.

\subsection{Preparing for the Tully-Fisher application} \label{tfsample} % 3.4    

In preparation for the cosmic flow fields analysis (see Sec.~1) we
determined which galaxies are potentially useful for inclusion in the
Tully-Fisher (TF) sample. Since the inclination corrections to the line
widths dominate the errors in the TF relation (note they are also used in
the internal absorption corrections in the TF relation, which however are
small in the NIR, see Table~\ref{tabext}), we selected only
`sufficiently' inclined galaxies for this sample, that is, galaxies with
minor/major axial ratios (2MASX parameter {\tt sup\_ba}) of $b/a \le 0.5$;
at relatively high extinction values, \ie $\ak \ge 0\fm50$, we also
selected galaxies with a generous $0.5 < b/a \le 0.7$ to account for the
fact that obscured galaxies appear rounder \citep{said15}. We thus have
selected 1296 galaxies from our ZoA sample and a further 35 galaxies from
the EBV sample (that is, 35\% and 40\% of the total samples, respectively).

Unlike the 2MTF survey we have not applied any further selection on galaxy
type. Although we could use the disc class parameter, a low surface
brightness disc can easily be missed in the ZoA. Instead, we flagged all
galaxies in this potential TF sample for \HI\ observations. Our final TF
sample to be used for cosmic flow studies will be based on their
\HI\ detection (see Paper II).

The photometry problems regarding the definition of the fiducial elliptical
aperture mentioned in Sec.~\ref{badphot} also affect the TF sample
definition: while most problems concern only the major axis position angle
(which can also affect the magnitude of a galaxy), in some cases the axial
ratio is affected and thus the derived inclination is incorrect. The effect
on inclination is small: based on visual inspection only an estimated
$2-3$\% of the total sample seems to have obviously incorrect inclinations
(see Item 4.\ in Sec.~\ref{badphot}). Note that for the final TF sample
selection and application all galaxies will be re-reduced and individually
inspected to ensure a high quality and consistency in all parameters (see,
for example, \citealt{said16b}). The extinction correction to the axial
ratios \citep{said15} will be applied at this stage leading to a final cut
in $b/a$, thus rendering the final TF sample complete and independent of
the above mentioned photometry problems.

\section{The catalogue} \label{results} % 4

The catalogue of the 6913 ZOA objects and the 502 EBV objects is available
online only as Tables 3a and 3b, respectively; an example page is given
below in Table~\ref{zoatabex}.

The columns are as follows:

\begin{description}

\item{\it Col. 1:} ID: 2MASX catalogue identification number (based on
  J2000.0 coordinates);

\item{\it Col. 2a and 2b:} Galactic coordinates: longitude $l$ and latitude
  $b$, in degrees;

\item{\it Col. 3:} Extinction: \ebv\ value derived from the DIRBE/IRAS maps
  (SFD98), in mag;

\item{\it Col. 4:} Object flag: 1 = obvious galaxy, 2 = galaxy, 3 =
  probable galaxy, 4 = possible galaxy, 5 = unknown, 6 = lower likelihood
  for galaxy, 7 = unlikely galaxy, 8 = no galaxy, 9 = obviously not a
  galaxy;

\item{\it Col. 5:} Object offset flag: `o' stands for coordinates that are
  offset from the centre of the object, and `e' stands for detections near
  the edge of an image (at an offset position from the object centre) which
  were detected with properly centred coordinates on the adjacent image;

\item{\it Col. 6:} Object class: `p' stands for PN, `p?' for possible PN,
  `s' for a probable DIRBE/IRAS point source that was not removed from the
  extinction maps (only the case if $|b|<5\degr$) and the extinction is
  likely to be overestimated, and `s?' stands for a possible DIRBE/IRAS
  point source (see Sec.~\ref{ext});

\item{\it Col. 7a:} Sample flag galaxy: `g' denotes a galaxy (object
  classes $1-4$), and `p' stands for galaxy candidates (class 5);

\item{\it Col. 7b:} Sample flag TF: `t' means the galaxy is inclined enough
  to be nominally included in the sample for application of the
  Tully-Fisher relation (see Sec.~\ref{tfsample});

\item{\it Col. 7c:} Sample flag \HI\ observation: `N' means the object was
  observed by us in the 21cm \HI\ line with the NRT (see Paper III), `P'
  means the object was observed by us with the Parkes radio telescope
  (publication in preparation), and a `+' indicates that the object
  still needs to be observed for a redshift;

\item{\it Col. 7d:} Sample flag extended sample: Objects that will {\it
  not} be included in the extended sample based on the diameter-extinction
  correction (\Kd\ $\le 11\fm25$) are marked with a star;

\item{\it Col. 8:} Galaxy disc type: `D1' means an obviously visible disc
  and/or spiral arms, `D2' stands for a noticeable disc, `D3' for a
  possible disc, and `D4' stands for no disc noticeable. Where it was not
  possible to tell (likely due to adjacent or superimposed stars or very
  high extinction) we give a flag `n';

\item{\it Col. 9:} 2MRS flags: `c' stands for an entry in the main
  catalogue, and `e' for the extra catalogue (that is, outside the 2MRS
  selection criteria) . For the sub-catalogues\footnote{for a detailed
    description see http://tdc-www.harvard.edu/2mrs/2mrs\_readme.html} we
  give: `cf' for an entry in the {\it flg} catalogue (affected by nearby
  stars, but not severely), `cp' for an entry in the {\it flr} catalogue
  (severely affected by nearby stars), `ca' and `ea' are objects that have
  been reprocessed and are both in the {\it rep} as well as {\it add}
  catalogues (in the latter case with new IDs and photometry, not listed by
  us), `cr' and `er' are objects rejected as galaxies and which can be
  found in the {\it rej} catalogues, `cn' and `en' are objects that have no
  redshifts yet and are listed in the {\it nocz} catalogues;

\item{\it Cols. 10a and 10b:} Velocity measurement flags `vo' and
  `vh' from the literature: `o' stands for an optical measurement and `o?'
  indicates a questionable optical measurement or object ID, `g' the
  velocity shows it to be a Galactic object, `h' stands for published
  \HI\ velocity and `h:' for uncertain \HI\ velocity, `h?' is an uncertain
  object ID (based on a radio telescope's large beam size that may include
  nearby galaxies), and a star `*' in either column stands for a yet
  unpublished measurement (see Sec.~\ref{vel});

\item{\it Col. 11:} Deep NIR image flag: `U' indicates there is a UKIDSS
  image for this object, and `V' stands for VISTA;

\item{\it Col. 12:} Photometry flag (Sec.~\ref{badphot}): the 2MRS flags
  for questionable photometry (see Col.~9) are repeated here: `a' stands
  for improved photometry exists, `p' for photometry is severely
  compromised, and `f' for poor photometry; in addition, `c' denotes
  improved photometry exists in John Huchra's original list, and `o' stands
  for offset (see Col.~4);

\item{\it Col. 13:} Extinction flag (see Sec.~\ref{badext}): `e' stands for
  a likely wrong extinction value, either due to small-scale variations or
  a non-removed IRAS/DIRBE point source, `e?' denotes a possible wrong
  extinction value;

\item{\it Col. 14:} 2MASX magnitude $K_{\mathrm 20}$: isophotal \K -band
  magnitude measured within the \K -band 20 mag arcsec$^{-2}$ isophotal
  elliptical aperture (in mag);

\item{\it Cols. 15 and 16:} 2MASX colours \hk\ and \jk : isophotal colours
  measured within the \K -band 20 mag arcsec$^{-2}$ isophotal elliptical
  aperture (in mag);

\item{\it Col. 17:} 2MASX flag $vc$: the visual verification score of a
  source (see Sec~\ref{colours});

\item{\it Col.18:} 2MASX object size $a$: the major diameter of the object,
  which is twice the 2MASX \K -band 20 mag arcsec$^{-2}$ isophotal
  elliptical aperture semi-major axis {\tt r\_k20fe} (in arcseconds);

\item{\it Col. 19:} 2MASX axis ratio $b/a$: minor-to-major axis ratio fit
  to the $3\sigma$ super-co-added isophote, {\tt sup\_ba}. Single-precision
  entries are given in cases where the 2MASX parameter {\tt sup\_ba} is not
  determined; these values were estimated from the axial ratios available
  in the different passbands;

\item{\it Col. 20:} 2MASX stellar density: co-added logarithm of the number
  of stars (\K$<14$\,mag) per square degree around the object;

\item{\it Cols. 21 -- 23:} Magnitude \Ko\ and colours \hko\ and
  \jko\ corrected for foreground extinction (Col.~3) as described in
  Sec.~\ref{extr} (in mag);

\item{\it Col. 24:} Extinction in the \K -band $\ak $: calculated from
  \ebv\ in Col.~4 and applying the SF11 correction of 0.86 (in mag);

\item{\it Col. 25:} Extinction corrected major diameter $a^{\rm d}$
  according to \citet{riad2010} using \ebv\ from Col.~4 with the SF11
  correction factor 0.86; only given for objects classified as galaxies (in
  arcseconds), see Sec.~\ref{diamcorr};

\item{\it Col. 26:} Extinction corrected magnitude \Kd\ corrected according
  to \citet{riad2010} using \ebv\ from Col.~4 with the SF11 correction
  factor 0.86; only given for objects classified as galaxies (in mag) (see
  Sec.~\ref{diamcorr});

\item{\it Cols. 27 and 28:} Extinction corrected colours \hkoc\ and
  \jkoc\ using \ebv\ from Col.~4 with the SF11 correction factor 0.86 (in
  mag).

\end{description}

\onecolumn 
{\tiny 
\input{2M_comb_ex.tex} 
} 
\twocolumn

\section{Properties of the catalogue } \label{prop}

\subsection{Distinguishing galaxies from non-galaxies } \label{colours}

\begin{figure} % 2
\centering
\includegraphics[width=0.45\textwidth]{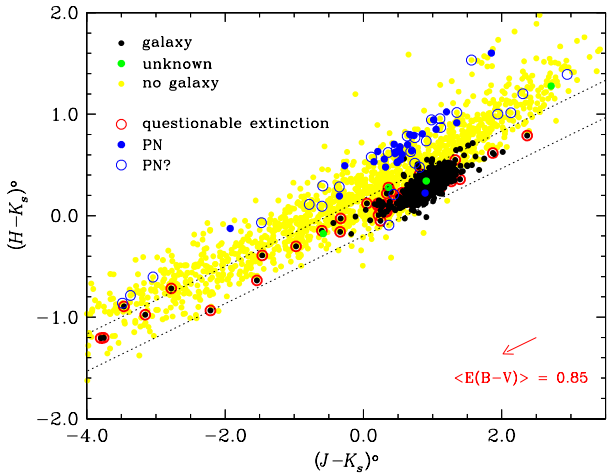}
\caption{2MASS extinction-corrected colour-colour plot: \hko\ as a function
  of \jko\ for all 2MASX objects in the ZOA and EBV samples. The reddening
  path for objects within the intrinsic colour range of galaxies is
  indicated with two parallel dashed lines. A reddening vector,
  representing the mean \ebv\ $=0\fm85$ of the sample, is shown as a red
  arrow in the bottom right corner. Black dots indicate galaxies, yellow
  dots non-galaxies, green dots objects of unknown nature, blue dots PNe
  and blue circles possible PNe; red circles denote objects with
  questionable Galactic extinction.
}
%\vspace{2cm}
\label{colcolplot}
\end{figure}

The cleanliness of the galaxy catalogue (\ie its contamination by
non-galaxies) can be visualised in a colour-colour plot such as
Fig.~\ref{colcolplot}. The colours are corrected for Galactic foreground
extinction according to SFD98 (see Sec.~\ref{sample}); please note that a
change in calibration in extinction as discussed in SF11 and Paper~III
introduces a stretching in colours but does not affect the qualitative
discussion presented here. Figure~\ref{colcolplot} shows objects classified
as galaxies in black and non-galaxies in yellow; the few galaxy candidates
(objects of unknown classification, class 5) are shown in green. Objects we
identified as Planetary Nebulae (PNe; based on information in the
literature or through visual inspection) are marked in blue (where open
blue circles denote possible PNe). They are clearly offset in
\hko\ colour. Also shown is the reddening path for the intrinsic colour
range occupied by the bulk of the galaxies (in-between the two dashed
lines). Galaxies with questionable extinctions (based on visually obvious
variations across the images), which are marked with large red circles, are
scattered along the reddening path. The plot suggests that all galaxies
with unusual \hko / \jko\ colour combinations in the reddening path have
incorrect extinction values attributed to them. However, galaxies outside
the reddening path are likely to be affected by starlight in the photometry
aperture. In other words, although we used the colour-colour plot during
the visual screening process to help distinguish some of the galaxy
candidates from other types of objects, this was not always possible, in
particular for objects within or near the reddening path (see, for example,
objects classified as `unknown').

\begin{figure} % 3
\centering
\includegraphics[width=0.45\textwidth]{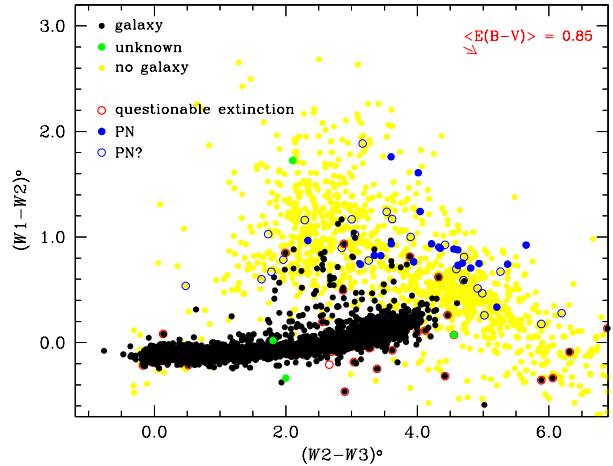}
\caption{WISE colour-colour plot: $(W1-W2)^o$ as a function of $(W2-W3)^o$ for
  all 2MASX objects in our ZOA and EBV samples. The symbols are as in
  Fig.~\ref{colcolplot}. 
}
\label{wisecolplot}
\end{figure}

We have made a similar colour-colour plot for the WISE bands $W1$ ($\lambda
= 3.4\mu$m), $W2$ ($\lambda = 4.6\mu$m) and $W3$ ($\lambda = 12\mu$m) using
aperture 1 ($5\farcs5$) data of the ALLWISE catalogue\footnote{We do not
  use total magnitudes; due to the large PSF these colours are likely
  contaminated by nearby stars, while colours in the NIR depend only very
  slightly on galaxian radius and are thus robust.}
Figure~\ref{wisecolplot} shows the extinction-corrected colour $(W1-W2)^o$
as a function of $(W2-W3)^o$ using the same symbols as in
Fig.~\ref{colcolplot}.  Using these colours, spiral galaxies and
ellipticals are separated in ($W2-W3$) (to the right and the left,
respectively, see Figure~26 in \citealt{jarrett11}), while active and
intensely star forming galaxies are separated in ($W1-W2$) (top). Unlike
for 2MASS colours, though, the PNe are not clearly offset from the
galaxies: their colours overlap with those of starburst or otherwise active
galaxies.

\begin{figure*} % 4
\centering
\includegraphics[height=0.85\textwidth,angle=270,origin=c]{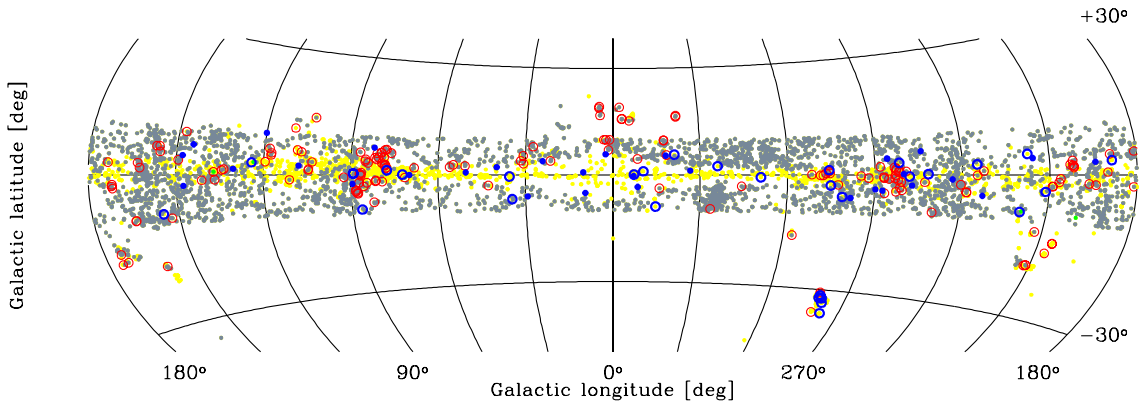}
\vspace{-3.5cm}
\caption{Aitoff projection in Galactic coordinates of the ZOA and EBV
  samples. The symbols used are basically the same as in
  Fig.~\ref{colcolplot} except that grey is used here for galaxies for
  better distinction.
}
\label{mapplot}
\end{figure*}

The on-sky distribution of both the ZOA and the EBV samples is shown in
Fig.~\ref{mapplot} with the basically same colour scheme as for
Figs~\ref{colcolplot} and \ref{wisecolplot}, except that galaxies are shown
in grey here for clarity. There are only a few areas at higher latitudes
($b\geq10$\degr) that have extinctions high enough to be in the EBV
sample. Most of the non-galaxies are close to the Galactic Plane which is
expected since the contamination of the 2MASX catalogue with blended stars
or with stars close to diffraction spikes of bright stars is small, and
most of the non-galaxian detections are caused by Galactic Nebulae.

The 2MASX catalogue gives a parameter {\it vc} which is a visual
verification score \citep{Jarrett+2000b}. It distinguishes between extended
sources ({\it vc} $=1$) and blended stars or artefacts ({\it vc} $=2$). In
quite a few cases the flags for `unknown' ({\it vc} $=-2$; $N=573$) and
`not examined' ({\it vc} $=-1$; $N=209$) are given. There is only one case
where we identified a {\it vc}-flag 2 object as a galaxy:
2MASX\,J18531497$-$0623155 is a tiny galaxy next to an equally bright star
and could only be unambiguously identified as a galaxy on a UKIDSS
image. This is a good example of a case where the superimposed star was not
subtracted from the photometry\footnote{The downloadable fits file
  obtainable at http://irsa.ipac.caltech.edu/applications/2MASS/PubGalPS/
  includes the star-subtracted images} and the galaxy itself is actually
too faint for our nominal magnitude limit.

\subsection{Completeness}   \label{compl} % 5.2

\begin{figure} % 5
\centering
\includegraphics[width=0.35\textwidth]{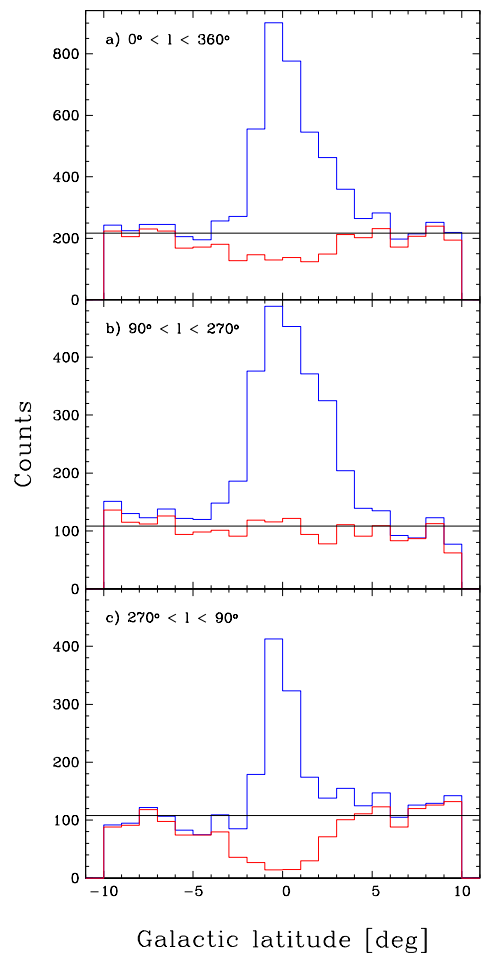}
\caption{Histograms of objects in the ZOA sample as a function of Galactic
  latitude for all objects (blue) and for galaxies only (red). The black
  line represents the average 2MASX bright galaxy count outside the ZoA.
  Panel (a): for all Galactic longitudes; panel (b) for the semi-circle
  towards the anti-centre region; panel (c) for the semi-circle towards the
  bulge.
}
\label{galbhist}
\end{figure}

\begin{figure*} % 6 
\centering
\includegraphics[width=0.95\textwidth]{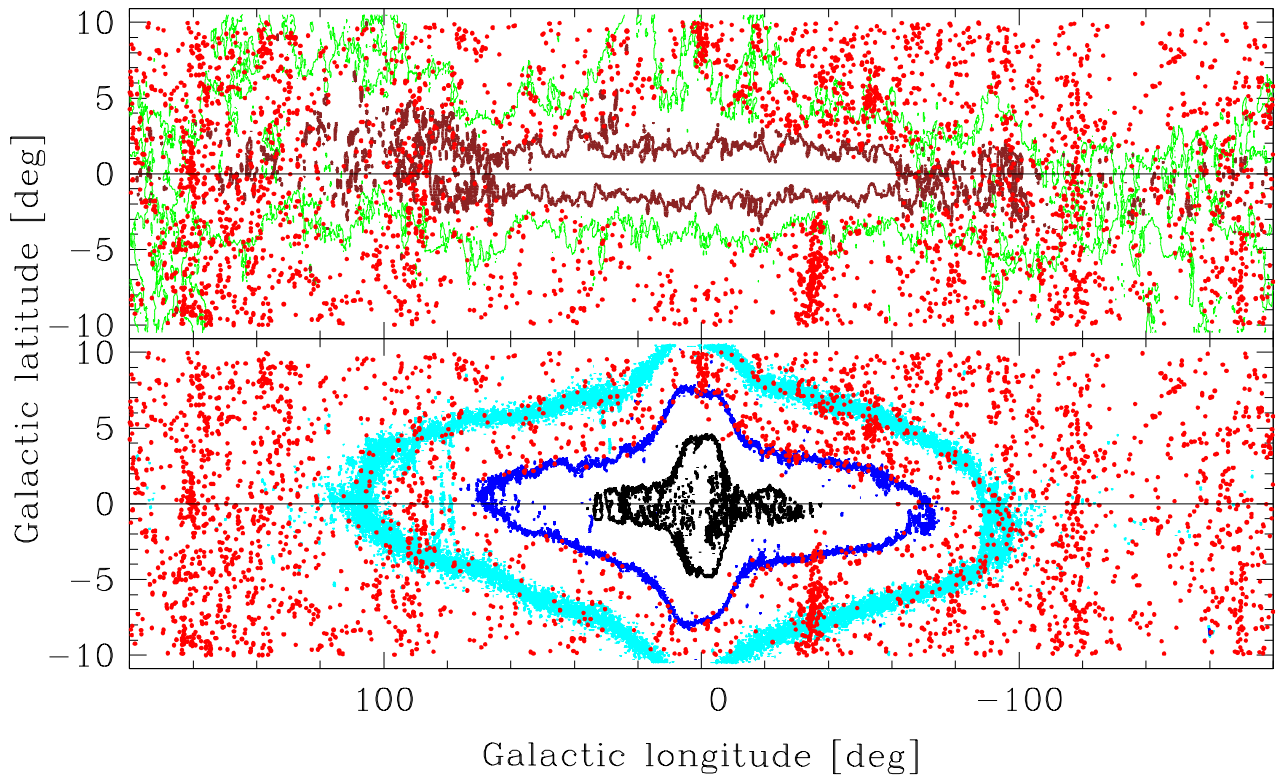}
\caption{Sky maps with contours of Galactic extinction (top) and stellar
  density (bottom) in the ZoA. Galaxies are indicated by red dots. Top
  panel: extinction contour levels shown are $\ak = 0\fm3$ (green) and
  $1\fm0$ (brown); bottom panel: stellar density contour levels shown are
  $\log N_*/{\rm deg}^2 = 4.0$ (light blue), 4.5 (dark blue) and 5.0
  (black).
}
\label{mapcontplot}
\end{figure*}

Most galaxy samples become incomplete at lower latitudes, and a lack of
galaxies in our sample in the Galactic bulge area is obvious from
Fig.~\ref{mapplot}. To test the completeness of the ZOA sample, we show in
Fig.~\ref{galbhist} (top panel) the histogram of source counts as a
function of Galactic latitude for all 6913 objects in our ZOA sample (blue)
and of the 3675 ZOA galaxies only (red). The black line represents the
average number of galaxies outside the ZoA of 0.605 galaxies/sq.deg, as
derived from the 2MASX catalogue for $|b|>10\degr$ and $K_{\rm 20} \le
11\fm25$, multiplied by the area covered in each bin. The histogram
for the galaxies shows a dip for $\sim-6\degr < b < 3\degr$, which means,
in this range we are incomplete. The dip disappears towards the anti-centre
region ($90\degr < l < 270\degr$, middle panel), while it is deeper for the
bulge region ($90\degr > l $ and $ l > 270\degr$, lower panel).

If we select only those latitudes bins with galaxy counts comparable to the
average number outside the ZoA (\ie $-10\degr < b < -6\degr$ and $3\degr <
b < 10\degr$), we find 2341 galaxies in our sample, while based on the
average 2MASX galaxy count outside the ZoA we would expect 2381
galaxies. The difference of 41 galaxies is smaller than the $1\sigma$
Poissonian error of 49 and therefore negligible. We thus conclude that, in
combination with the all-sky 2MASX sample, we are now complete below
$b=-6\degr$ and above $b=3\degr$, and the 2MASX bright galaxy ZoA is
reduced to a strip of 9\degr width in Galactic latitude. In the anti-centre
region the ZoA has all but disappeared.

We also note a slight slope with latitude in the lower two panels: while
towards the anti-centre region the area above the Galactic plane shows
fewer galaxies than below the plane (345 and 489 for the four outermost
bins, respectively), the trend is reversed for the Galactic bulge region:
466 and 395, respectively. This asymmetry seems to be entirely due to large
scale structures and varies depending on where the longitude cuts are set.

Both effects are also evident in the on-sky distribution of the ZOA
galaxies as shown in Fig.~\ref{mapcontplot} (red dots): the top panel
displays contours of Galactic extinction, and the bottom panel shows
contours of stellar density as derived from the 2MASS Point Source
Catalogue\footnote{Using the same definition as for the stellar density
  parameter in the 2MASX catalogue, that is, the $\log$[number of stars
  deg$^{-2}$] of stars with \K$<14$\,mag.}. While the distribution of
galaxies is influenced by density variations due to obvious large scale
structures, it is evident that the lack of 2MASX galaxy detections towards
the Galactic bulge region is mainly due to the high stellar density: we
lose completeness at around or slightly above the $\log N_*/{\rm deg}^2$ =
4.5 level (dark blue contour); the extinction contours cannot explain in
particular the lack of 2MASX galaxies at the higher latitudes around the
Galactic Centre. On the other hand, even at extinctions as high as $\ak =
1\fm0$ (brown contour) it is possible to detect 2MASX galaxies. Note also
the slight tilt angle in the contours of both extinction and stellar
densities which is due to the Galactic warp (\eg \citealt{reyle09}).

\begin{figure} % 7 
\centering \includegraphics[width=0.40\textwidth]{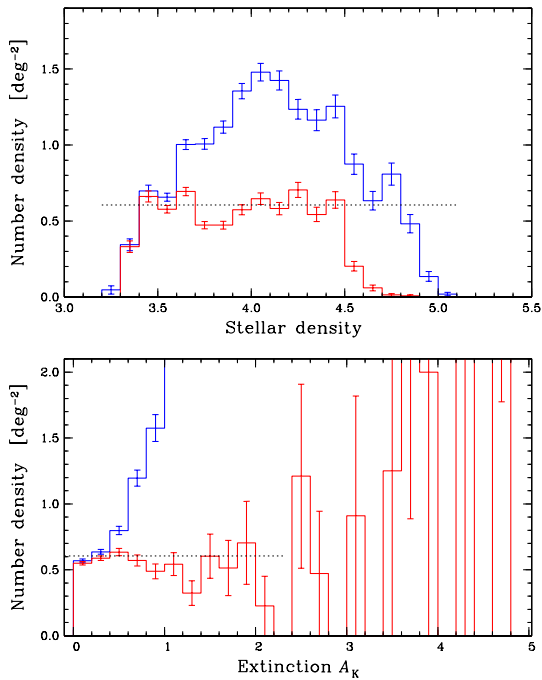}
\caption{Histograms of the number density of all ZOA sample objects (blue)
  and galaxies only (red) as a function of stellar density (top panel) and
  Galactic extinction $\ak $ (bottom panel; only regions with $\log
  N_*/{\rm deg}^2 \le 4.5$ were used). The error bars indicate Poissonian
  errors per bin. The black dotted lines represent the average 2MASX
  bright galaxy number density outside of the ZoA.
}
\label{histboth}
\end{figure}

For a more quantitative analysis, we can use histograms of galaxy counts as
a function of extinction and stellar density. A comparison is only useful,
however, if we take into account the variation in sky area per bin. We have
thus binned the extinction and stellar density data in the ZoA (excluding
the EBV sample) and derived number density values (per sq.deg) as shown in
Fig.~\ref{histboth} (red for galaxies, blue for all objects).

The number density of galaxies remains remarkably constant across all
stellar densities in the ZoA except for the lowest and highest values (top
panel): for $\log N_*/{\rm deg}^2 <3.4$ the overall area is very small
(3\%), whereas the abrupt cut-off at $\log N_*/{\rm deg}^2 >4.5$ is due to
incompleteness (see also \citealt{Jarrett+2000b}). The bin-to-bin
variations in the range $3.4 - 4.5$ are most likely due to variations in
large-scale structures. The histogram also shows that the non-galaxy
population (the difference between the blue and red histograms) increases
rapidly with stellar density up to $\log N_*/{\rm deg}^2 \sim 4.0$ and
accounts for most 2MASX detections at $\log N_*/{\rm deg}^2 >4.5$ (87\%).

To investigate whether extinction has any effect on the completeness we
first excluded all areas with stellar densities $\log N_*/{\rm deg}^2 >4.5$
(to avoid a selection bias since both high extinction and high stellar
density occur in similar areas, see Fig.~\ref{mapcontplot}).  The resulting
histogram is shown in the bottom panel of Fig.~\ref{histboth}. The number
densities are highly affected by low number statistics for $\ak \,
\approxgt \, 1\fm0$ (88 galaxies). All the bins up to $\ak =2\fm0$ are
consistent within the errors with the high-latitude 2MASX galaxy density of
0.605 galaxies/sq.deg (dotted line). The slight apparent drop off from $\ak
\sim 0\fm6$ towards $ 1\fm4$ is not significant and could also be due to
variations in large scale structures.

Our conclusion is that the 2MASX sample of \Ko\ $\le 11\fm25$ is complete
over the whole sky wherever $\log N_*/{\rm deg}^2 <4.5$ and at least where
$\ak < 0\fm6$ (or $E(B-V)=1\fm61$), and likely up to $\ak \sim 2\fm0$ (or
$E(B-V)=5\fm37$). Thus the NIR `bright galaxies' ZoA has been reduced to
cover only 2.4\% of the whole sky as compared to the over 20\% for the
original optical ZoA (based on a diameter limit of $1\arcmin$,
  see \citealt{KKLahav2000})  and 9.6\% for 2MRS.

\subsection{Extinction} \label{ext} % 5.3

We have set two flags in the catalogues regarding potential extinction
problems: (i) for objects lying in areas where we suspect that the
extinction varies rapidly we have set an extinction flag (Col.~13; see also
Figs~\ref{colcolplot}\,--\,\ref{mapplot}); (ii) for objects that seem to be
either a source in the DIRBE/IRAS FIR maps or affected by one, we have set
an object class (Col.~6) which was determined as follows.

In the SFD98 maps, FIR sources (mostly extragalactic, but also unresolved
Galactic sources) were removed for $|b|>5\degr$. At lower latitudes,
sources were only removed from some selected (unconfused) areas. The output
of the query script for a given position gives flags whether a source list
exists, and if a source has been subtracted. For 3184 objects and 2476
galaxies a source list was available, with a source being subtracted in 702
and 541 cases, respectively.

\begin{figure} 
\centering
\includegraphics[width=0.35\textwidth]{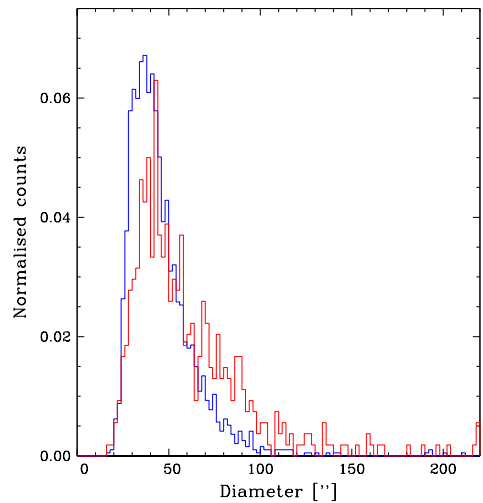} % 8
\caption{Normalised histogram of major axis diameters of galaxies (in
  arcseconds) for those galaxies where a DIRBE/IRAS FIR source list was
  used in the making of the SFD98 Galactic extinction maps. Red: all
  positions where a source was subtracted, blue: all other positions.
}
\label{drbhist}
\end{figure}

Next, we investigated the SFD98 maps of all our galaxies for which no
DIRBE/IRAS source list exists ($N=1287$) to establish whether the position
might be affected by an unsubtracted point source. We compared the
extinction value at the target position to those measured at four positions
offset by seven arcminutes. Furthermore we investigated all galaxies with a
diameter larger than 100 arcseconds since these are more likely bright in
the FIR, see Fig.~\ref{drbhist}: the blue histogram show the major axis
diameters of all galaxies that lie in areas where a DIRBE/IRAS source list
exists, and the red histogram is for objects where a source was
subtracted. The figure shows that larger galaxies are more likely to be
detected in the FIR and thus have their FIR emission subtracted from the
Galactic extinction maps.

For any suspicious case we checked whether the source was detected by
IRAS. We have flagged five galaxies as DIRBE/IRAS sources or being close to
one: the three large, nearby galaxies Circinus, Maffei 2 and IC\,10 seem to
be sources themselves, while 2MASX\,J19241426+2047311 and
2MASX\,J06142070+1349298 are close to compact \HII\ regions. We identified
a further six galaxies which seem to be associated with weak FIR sources.

While we identified $<1$\% of low-latitude galaxies to be obviously
affected by FIR sources, we expect more than a fifth of the low-latitude
galaxies (or $\sim\!250$) to be affected by weaker FIR sources, based on
the 541 galaxies (or 22\%) that are affected in the area where a DIRBE/IRAS
source list was used to determine the Galactic extinction.  We do not have
the proper means however to make a comprehensive census of all these
cases. For our current purpose it is sufficient to flag the most severe
cases and to emphasise that weak unsubtracted FIR sources at low latitudes
are likely to affect the derived extinction, similar to the spatial
variability cases. This will be particularly important for any flow field
analysis.

\subsection{Galaxy types} \label{disctype}  % 5.4

Morphological classification is difficult in the NIR, and with the
foreground extinction mainly affecting the outer discs such that the
bulge-to-disc ratio becomes uncertain, we have not attempted to do this for
our sample. Since one of the goals of the project is to obtain
\HI\ observations for a TF analysis, we classified our galaxies on the
presence or absence of a disc component. We give four disc classes
D1\,--\,D4, ranging from obvious disc visible (often spiral arms are also
discernible) to no visible disc. In cases where it was not possible to
distinguish a disc component, \eg in very high extinction areas or where
stars obscure most of the galaxy, we have set a flag `n'.

\begin{table}  % 4
\centering
\begin{minipage}{140mm}
\caption{{Disc classification of galaxies in both samples } \label{tabdisc}}
\begin{tabular}{llrr}
\hline
Class & meaning     & ZOA sample & EBV sample \\
\hline
D1  & obvious disc   & 2085 &  41 \\  %class=1
D2  & disc           &  634 &  29 \\  %class=2
D3  & possible disc  &  538 &  11 \\  %class=3
D4  & no disc        &  319 &   5 \\  %class=4,5
n   & unable to tell &   99 &   2 \\  %class=9
\hline
total &                & 3675 &  88 \\
\hline
\end{tabular}
\end{minipage}
\end{table}

Table~\ref{tabdisc} gives the disc class statistics for the ZOA and EBV
samples. We find that 76\% of all galaxies show a disc, while 9\% show no
disc (we exclude the non-classifiable cases here). Considering that the
global E : S galaxy morphology ratio is roughly 20\% : 80\% suggests
that the D3 class (15\%) is mostly comprised of elliptical
galaxies. Because the presence of a disc is easier to establish than its
absence, classes D1 and D2 are more reliable than, \eg class D4.

\begin{figure*} % 9
\centering
\includegraphics[width=0.95\textwidth]{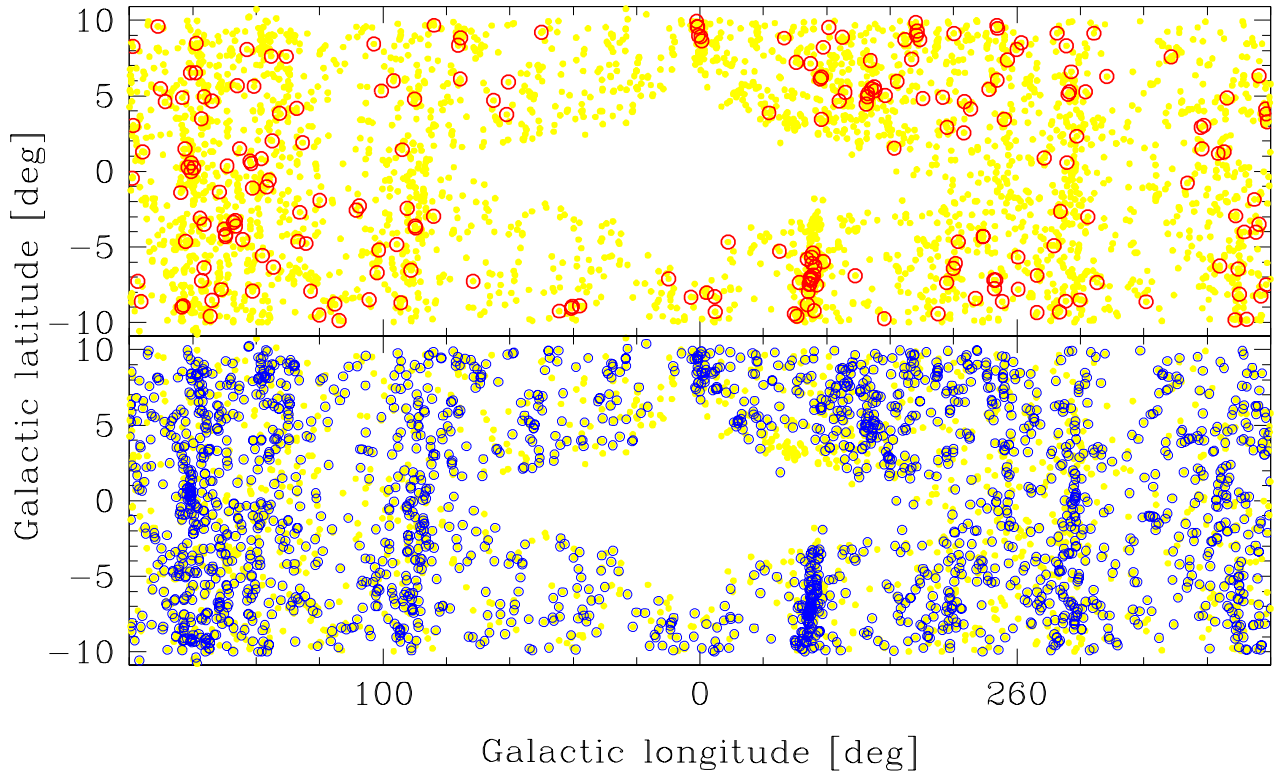}
\caption{Comparison of the sky distributions of ZOA galaxies with and
  without a discernible disc in the NIR. All galaxies are indicated by
  yellow dots. Top panel: Galaxies with a D4 classification (\ie no visible
  disc) are shown as red circles. Bottom panel: Galaxies with a D1
  classification (\ie with a clear disc) are shown as blue circles.
}
\label{mapclplot}
\end{figure*}

\begin{figure*} % 10
\centering
\includegraphics[width=0.95\textwidth]{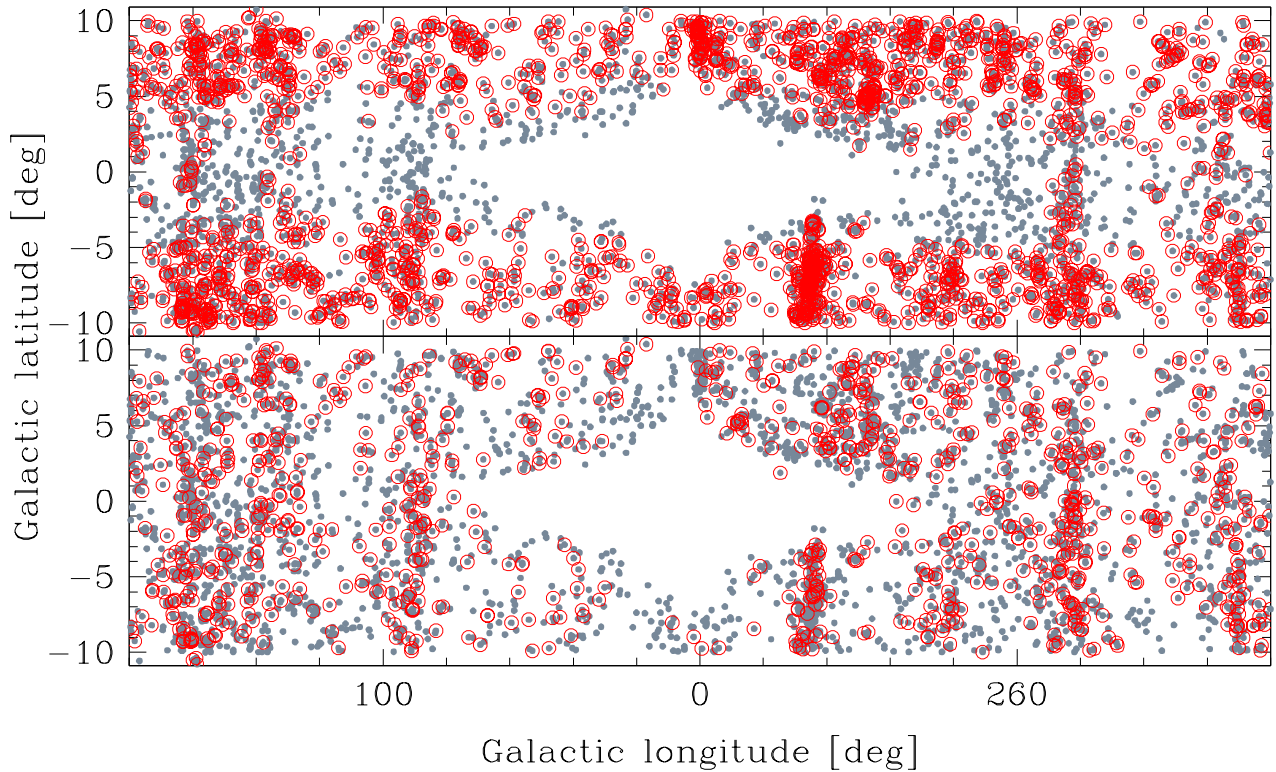}
\caption{Sky distributions of ZOA galaxies with or without known
  redshifts. All galaxies are indicated by grey dots; red circles indicate
  objects with redshift measurements. The top panel shows optical redshift
  measurements ($N=2241$), the bottom panel \HI\ detections ($N=963$).
}
\label{mapvelplot}
\end{figure*}

In Fig.~\ref{mapclplot} we show the sky distribution of the two extreme
classifications, $D4$ (no disc, red circles in top panel) and $D1$ (clear
disc, blue circles in bottom panel). In the case of $D4$ (which basically
stands for elliptical galaxies) we would expect to see the cores of rich
clusters, and indeed we can discern, for example, the Norma cluster (Abell
3627, $l\simeq325\degr$, $b\simeq-7\degr$), the Ophiuchus Cluster
($l\simeq0\degr$, $b\simeq9\degr$) and the cluster around 3C129
($l\simeq160\degr$, $b\simeq0\fdg5$) which was revealed to be a rich
cluster through our NRT observations (Paper II) and the follow-up
observations with the Westerbork Synthesis Radio Telescope
\citep[][]{ramatsoku14, ramatsoku16}.

As there are about nine times more galaxies classified as $D1$ (clear disc,
that is, spiral galaxies) the bottom panel is more crowded, revealing more
filamentary structures, predominantly in the N-S direction: \eg the Puppis
filament at $l\simeq245\degr$, as well as the western and eastern arm of
the Perseus-Pisces filament ($l\simeq160\degr$ and $l\simeq90\degr$,
respectively; see \citealt{ramatsoku14}, Paper II).

\subsection{Radial velocity information } % 5.5

For 2620 galaxies (70\%; with 2549 in the ZOA sample) we have found at
least one radial velocity measurement in the literature. For a further 293
(279) galaxies, redshifts will be publicly available presently:
six are optical measurements (from the updated 2MRS catalogue); the
\HI\ measurements are mainly from our observing campaigns at \nan\ and
Parkes (see Sec.~\ref{vel} for details). This will increase the total
number of galaxies with at least one redshift measurement to 2805
(2729). All redshifts will be presented in a forthcoming paper in
combination with an analysis of the whole-sky redshift distribution.

When we plot the distribution of these galaxies on the sky and distinguish
between optical and \HI\ redshifts (see the red circles in the upper and
lower panels, respectively, of Fig.~\ref{mapvelplot}), two features are
immediately obvious: (i) the inner ZoA is dominated by \HI\ detections, and
(ii) the \HI\ detections, unlike the optical detections, do not show
pronounced clustering \citep{Koribalski2004}. Both effects are typical for
the respective types of measurement.

\begin{figure} % 11
\centering
\includegraphics[width=0.35\textwidth]{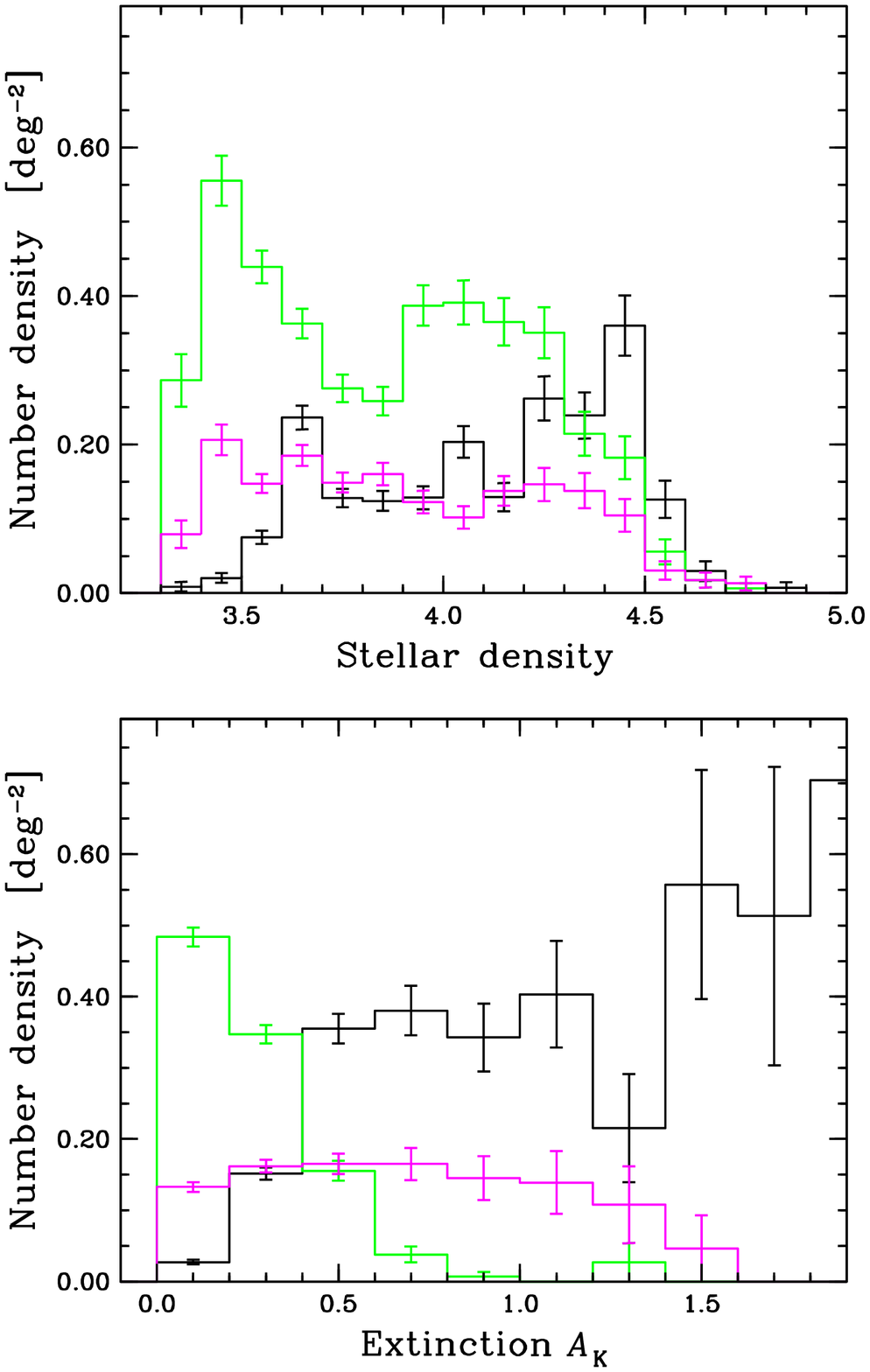}
\caption{Histograms of the number density of ZOA sample galaxies with
  optical redshifts (green), \HI\ redshifts (magenta) and no velocity
  information at all (black). Top panel: as a function of stellar density;
  bottom panel: as a function of Galactic $\ak $ (only regions with $\log
  N_*/{\rm deg}^2 \le 4.5$ were used). Poissonian errors per bin are
  indicated with error bars.
}
\label{velhist}
\end{figure}

We show the number densities of galaxies in the ZOA sample with and without
velocity information as a function of stellar density and extinction $\ak $
in Fig.~\ref{velhist} (top and bottom panel, respectively) using the same
method as for Fig.~\ref{histboth}. As before, the extinction histogram has
been restricted to areas with stellar densities $\log N_*/{\rm deg}^2 <
4.5$. Though we would expect the number density of galaxies with
\HI\ measurements (magenta; $N=963$) not to show any dependence on either
stellar density or extinction, we find a slight drop-off in both
histograms. This is understandable since most \HI\ observations in the
north are from pointed observations of recognisable galaxies in
lower-stellar density environments; only one (shallow) blind survey exists
so far, \ie EBHIS \citep{kerp11}. On the other hand, number densities of
galaxies with optical measurements (green; $N=2241$) show, as expected, a
strong decrease with extinction, but also a dependence on stellar
densities. This is partially due to the fact that many of the optical
measurements have been obtained by the 2MRS collaboration and thus are
restricted to higher latitudes where both extinction and stellar densities
are lower. Note that for 475 galaxies both kinds of velocity measurements
are available. For completeness, we also show galaxies without any velocity
information (black; $N=946$). 

\subsection{2MASS parameters }  \label{2mparam}  % 5.6 

Throughout the following analysis we have excluded all galaxies which have
a photometry or extinction flag set, or for which $ \ak > 3\fm0$. Unless
stated otherwise, we used the combined ZOA \& EBV catalogue. Histograms
were normalised by total number in the regarded sample.

\subsubsection{Diameters and magnitudes } 

\begin{figure} % 12
\centering
\includegraphics[width=0.35\textwidth]{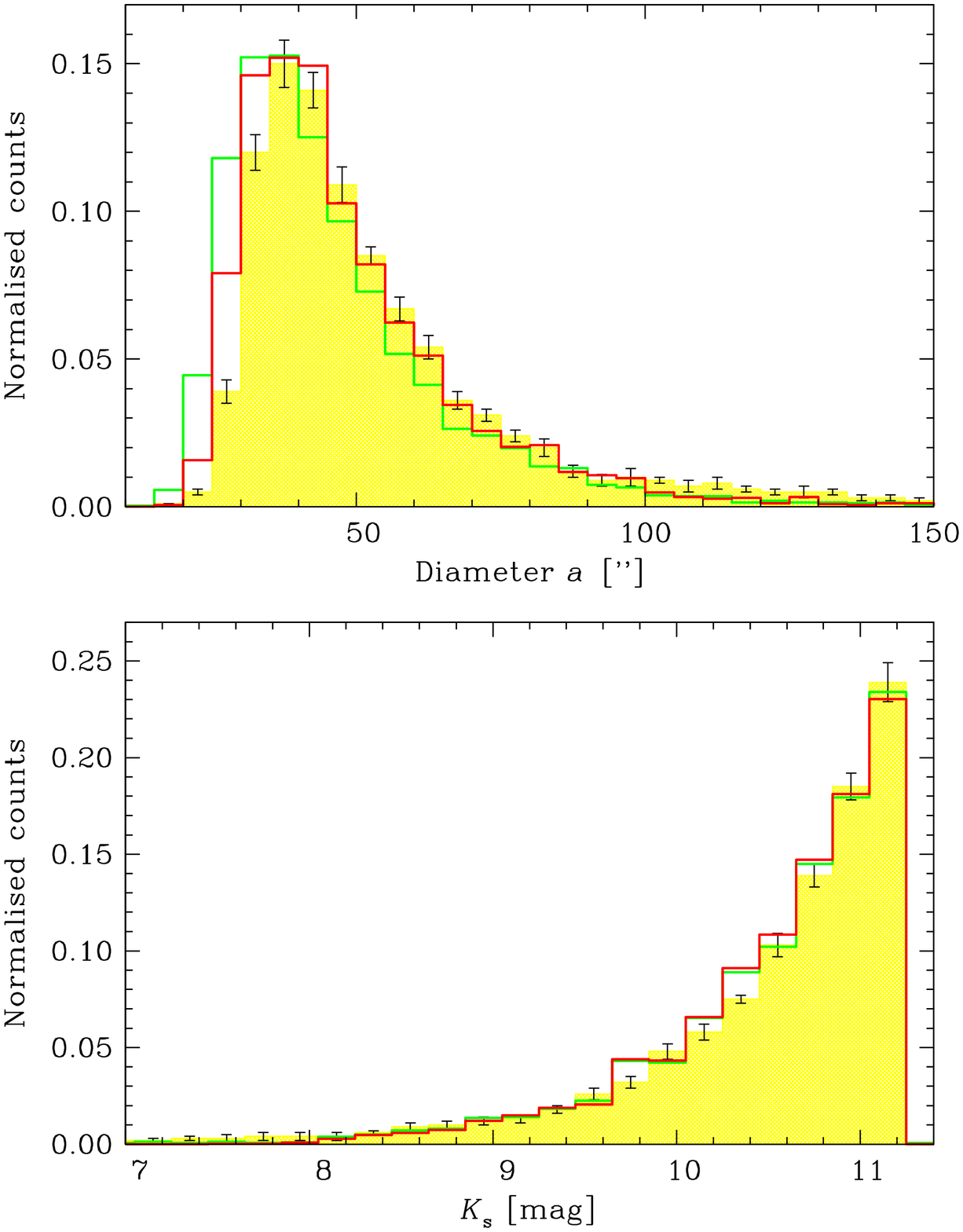}
\caption{Normalised histograms of major axis diameters in arcseconds (top)
  and \K -band magnitudes (bottom panel) for 2MASX galaxies. Yellow
  histograms represent the average high Galactic latitude sample; green
  histograms are for the combined ZOA \& EBV galaxy sample (where the \K-
  band magnitude is corrected for Galactic foreground extinction, \Ko );
  red histograms are the same but with the additional, diameter-dependent
  extinction applied (\ad\ and \Kd ).
}
\label{2mhist}
\end{figure}

Figure~\ref{2mhist} shows the normalised histograms of the parameters major
axis diameter $a$ (top panel) and \K -band magnitude (bottom panel) for
various samples. First, we compare the parameters of our sample (green
lines, with magnitudes corrected for foreground extinction, \ie \Ko ), with
those corrected for the diameter-dependent extinction (see
Sec.~\ref{diamcorr}), \ad\ and \Kd\ (red lines). We find that while the
diameters \ad\ show a clear shift to larger values compared to $a$ (as
expected), there is no obvious similar shift towards brighter \K -band
magnitudes.

\begin{table}
\centering
\begin{minipage}{140mm}
\caption{{Statistics on 2MASX parameters of various samples} \label{tab2mhist}}
{\scriptsize
\begin{tabular}{l@{\extracolsep{0.0mm}}c@{\extracolsep{2.0mm}}r@{\extracolsep{3.0mm}}r@{\extracolsep{2.0mm}}c@{\extracolsep{2.0mm}}c}
\hline
Parameter (sample) & $N^*$ & min & max & median & error \\
\hline
\multicolumn{6}{l}{Diameter:} \\ 
$a$ (comb${\dagger}$)     & 3365 & $15\farcs0$ &   573\arcsec   &  $41\farcs2$ & 0.5 \\
\ad (comb${\dagger}$)     & 3301 & $17\farcs5$ &   273\arcsec   &  $43\farcs6$ & 0.4 \\
$a$ (high-lat$^{\ddagger}$) &$2950-3374$& $13\farcs6$ &  1260\arcsec &  $45\farcs8 - 49\farcs6$ & $0.4-0.9$ \\
\ad\ (ext'd$^{\rm \mathsection}$)   & 3344 & $17\farcs3$ &   273\arcsec   &  $43\farcs4$ & 0.4 \\
\hline
\multicolumn{6}{l}{\K -band magnitude:} \\ 
\Ko (comb${\dagger}$)     & 3365 & $4\fm32$    &  $11\fm25$       &  $10\fm74$    & 0.01 \\
\Kd (comb${\dagger}$)     & 3301 & $6\fm78$    &  $11\fm25$       &  $10\fm74$    & 0.01 \\
\K\ (high-lat$^{\ddagger}$) &$2950-3374$& $1\fm55$    &  $11\fm25$  &  $10\fm72 - 10\fm78$    & $0.01-0.02$ \\
\Kd\ (ext'd$^{\rm \mathsection}$)   & 3344 & $6\fm78$    &  $11\fm25$       &  $10\fm75$    & 0.01 \\
\hline
\multicolumn{6}{p{0.55\textwidth}}{$^*$ sample size; $^{\dagger}$ combined ZOA and EBV sample; \,\,\, \,\,\, \,\,\, $^{\ddagger}$
  high-latitude sample; $^{\rm \mathsection}$ extended sample} \\
\end{tabular}
}
\end{minipage}
\end{table}

To assess whether the diameter-dependent extinction correction is adequate,
we compared our sample parameters with those of galaxies well outside the
ZoA. We extracted two samples of 2MASX galaxies from high latitudes ($|b| >
30\degr$) with the same magnitude limit of \K $<11\fm25$ (uncorrected for
foreground extinction). The samples contain $6747$ objects from the
northern Galactic cap and $5900$ from the southern Galactic cap.  As these
numbers are affected by cosmic variance, we used a simplistic method to
represent this effect by extracting 12 subsamples from the total sample,
each with $\sim\!3000$ galaxies selected using different criteria (either
random or by sorting in RA).  We then took the mean of these normalised
histograms and determined the scatter in each bin. The result is shown as
the yellow filled (averaged) histograms in Fig.~\ref{2mhist}, where the
black error bars represent the scatter in those bins.

To quantify the results, Table~\ref{tab2mhist} gives sample size $N$, the
minimum and maximum\footnote{The largest extinction-corrected diameter
  \ad\ is smaller than the largest uncorrected diameter because the largest
  galaxy has no corrected parameter, since for some cases the surface
  brightness parameter necessary for the correction was not specified in
  the 2MASX catalogue.}  values as well as median values for the different
samples. We decided to use the median values for the comparison since the
histograms are skewed, and we cannot apply the KS-test. As an estimate for
the uncertainties we list the errors in the mean (which compare reasonably
well with the dispersion in the median of the high-latitude samples).

As already noticed in Fig.~\ref{2mhist}, the median value of the diameters
of our sample shows a clear shift from uncorrected, $a$, to corrected
diameters, \ad , by $\sim\!6\sigma$. However, all high-latitude samples
have diameters even larger by at least $5\sigma$ compared to our \ad , and
by $8\sigma$ compared to the median value of the full high-latitude sample
($N=12,647$), $47\farcs0$. This suggests that the diameter-dependent
extinction correction as derived by \citet{riad2010} is too conservative
and not quite sufficient for our sample. A selection bias with respect to
size as an alternative explanation is unlikely because we select on
magnitudes which show no difference (see discussion below).

In case of the \K -band magnitudes, in the ZoA the median \Kd\ magnitude is
the same as the median \Ko\ magnitude, and both lie well within the range
of median values found for the high-latitude samples. It is important to
note, though, that with the diameter-dependent correction our sample
cut-off at \Ko $= 11\fm25$ no longer results in a complete sample since we
lack those galaxies which are fainter than \Ko\ but have \Kd
$\le11\fm25$. We therefore extended our sample by (a) adding the
supplementary \Kd -sample described in the appendix (with a new magnitude
cut at \Kd $= 11\fm25$; see also Sec.~\ref{diamcorr}) and (b) excluding
other galaxies where the reduced extinction correction due to the SF11
factor 0.86 makes them fainter than our magnitude limit (see appendix) so
that we again have a complete magnitude-limited sample (called the extended
sample in Table~\ref{tab2mhist}). We find that the median \Kd\ magnitude of
$10\fm75$ for this extended sample is comparable to the previous
values. This is expected because there are two, opposing, corrections which
almost cancel each other out: the diameter-dependent extinction correction
makes magnitudes brighter, whereas the correction factor $f=0.86$ (SF11),
which is also applied when deriving \Kd\ values, makes magnitudes fainter.

On the other hand, the median extinction corrected diameter \ad\ of this
extended sample is slightly smaller than previously and compared to the
high-latitude samples it is significantly smaller by $\sim\!9\sigma$.

As a sanity check on whether the enforced cut-off has an influence on the
median we did the same tests on a diameter-limited subsample using a lower
diameter limit of 40 arcseconds. The results confirmed the above findings.

It thus seems that the diameter-dependent extinction correction for the
diameter is not sufficient and that intrinsically the galaxies in the ZoA
are likely to be larger (by $\sim\!4\arcsec$).  The diameter-dependent
extinction correction for the \K -band magnitudes is not affected by this
underestimation since the magnitude correction is independent of the
actual diameter of a galaxy \citep{Cameron1990}.

\subsubsection{NIR colours}  % 5.6.2

\begin{figure} % 13
\centering
\includegraphics[width=0.35\textwidth]{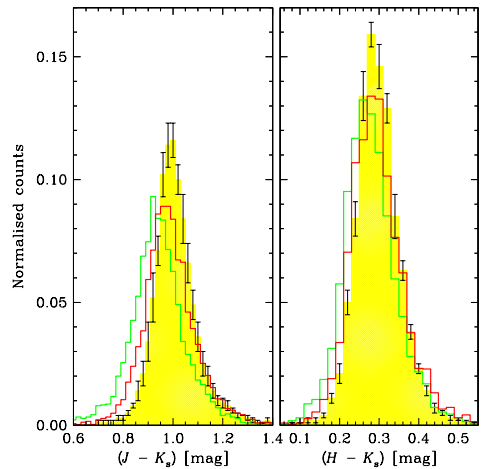}
\caption{Normalised histograms of 2MASX colours \jk\ (left) and
  \hk\ (right). Yellow histograms represent the average high Galactic
  latitude sample; green histograms are the combined ZOA \& EBV galaxy
  sample (corrected for Galactic foreground extinction); red histograms are
  the latter sample but with the SF11 correction factor 0.86 used in the
  extinction correction.
}
\label{2mhist2}
\end{figure}

The normalised histograms for the colours \jk\ and \hk\ are shown in
Fig.~\ref{2mhist2}. As for Fig.~\ref{2mhist}, the green histograms show the
extinction-corrected colours (\hko\ and \jko ) of our sample, while the
yellow filled histograms are the average of the high-latitude samples
(where no extinction correction has been applied).  As explained in
Sec.~\ref{diamcorr}, the colours are not affected by the diameter-dependent
extinction correction though we did apply the correction factor 0.86 (SF11)
to the \ebv\ values used in the extinction correction; these colours are
shown as red histograms (designated as \hkoc\ and \jkoc ).

Although the scatter in the colours is large since they are sensitive to
problems with extinction and photometry, the mean values are well defined:
$<(J-K_{\rm s})^{\rm o}> = 0\fm938 \pm 0\fm002$ and $<(H-K_{\rm s})^{\rm
  o}> = 0\fm279 \pm 0\fm002$ with a standard deviation of $\pm$$0\fm13$ and
$\pm$$0\fm08$, respectively. These values are both bluer than those of the
full high-latitude sample ($1\fm013 \pm 0\fm001$ and $0\fm302 \pm 0\fm001$,
respectively). The application of the improved extinction correction as
recommended by SF11, however, changes the mean colours to $0\fm988$ and
$0\fm296$, respectively, which is in good agreement with the high-latitude
samples. We summarise the statistics of the various samples in
Table~\ref{tab2mhistb} where we give the sample size $N$, the mean and its
error, as well as the standard deviation (or scatter) of the samples.

\citet{Jarrett2000} and \citet{jarrett03} present NIR colours as a function
of morphological type: for early-type spirals they find a \jk\ value
comparable to ours, that is, $\sim\!1\fm0$, while their \hk\ colour is
bluer: $0\fm27$. Later type spirals are generally bluer. The comparison,
though, is affected on the one hand by the small sample size in the large
galaxy catalogue (\citealt{Jarrett2000}, where the total sample consists of
100 galaxies), and on the other hand by our sample including galaxies at
larger redshifts which appear redder (though the contamination by these is
constrained by our bright magnitude cut-off).

\begin{table*}
\centering
\begin{minipage}{280mm}
\caption{{Statistics on 2MASX colours of various samples} \label{tab2mhistb}}
{\scriptsize
\begin{tabular}{l@{\extracolsep{0.0mm}}c@{\extracolsep{2.0mm}}c@{\extracolsep{2.0mm}}c@{\extracolsep{0.0mm}}c@{\extracolsep{2.0mm}}c@{\extracolsep{2.0mm}}c@{\extracolsep{2.0mm}}c@{\extracolsep{0.0mm}}c}
\hline
Sample       & \multicolumn{4}{c}{\jk } & \multicolumn{4}{c}{\hk } \\
\noalign{\smallskip}
\cline{2-5} \cline{6-9} 

\noalign{\smallskip}
       & $N^*$ & mean & error & standard deviation & $N^*$ & mean & error & standard deviation \\
       &     & [mag]& [mag] & [mag]              &     & [mag]& [mag] & [mag]              \\
\hline
Colour$^{\rm o}$\, (comb$^{\dagger}$)   & 3255 & $0.938$ & $0.002$ & $0.13$ & 3322 & $0.279$ & $0.001$ & $0.08$ \\
Colour$^{\rm o,c}$\, (comb$^{\dagger}$) & 3255 & $0.988$ & $0.002$ & $0.12$ & 3322 & $0.296$ & $0.001$ & $0.07$ \\
Colour$$\, (high-lat$^{\ddagger}$)       &$2939-3372$ & $1.000-1.022$ & $0.002-0.003$ & $0.10-0.19$ & $2939-3372$ & $0.294-0.313$ & $0.001-0.002$ & $0.07-0.13$ \\
Colour$^{\rm o}$\, (comb, corr$^{\rm \mathsection}$)   & 3220 & $0.962$ & $0.002$ & $0.13$ & 3287 & $0.293$ & $0.001$ & $0.08$ \\
Colour$^{\rm o,c}$\, (comb, corr$^{\rm \mathsection}$) & 3220 & $1.012$ & $0.002$ & $0.12$ & 3287 & $0.310$ & $0.001$ & $0.08$ \\
\hline
\multicolumn{9}{p{17cm}}{{$^*$ sample size; $^{\dagger}$ combined ZOA and EBV sample; $^{\ddagger}$
  high-latitude sample; $^{\rm \mathsection}$ combined ZOA and EBV sample corrected for
  stellar density dependence}} \\
\end{tabular}
}
\end{minipage}
\end{table*}

\begin{figure} % 14
\centering
\includegraphics[width=0.45\textwidth]{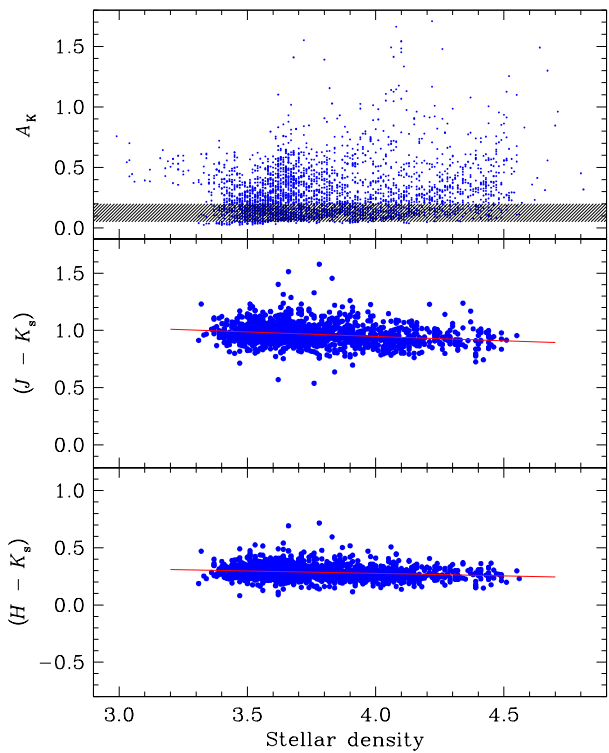}
\caption{Top panel: Galactic extinction $\ak $ as a function of stellar
  density for our galaxy sample. The shaded area indicates the range of
  $0\fm05< A_{\rm K} < 0\fm2$ where the stellar density and extinction are
  not correlated (see text). Middle panel: colour \jko\ as a function of
  stellar density, for the sample defined by the shaded area in the top
  panel. Bottom panel: same for the \hko\ colour. The red lines show the
  linear fits to the distributions.
}
\label{colstdens}
\end{figure}

As we have shown above, some properties of ZoA galaxies are not only
affected by Galactic foreground extinction but also by stellar
density. Since there is a large overlap between regions with high stellar
densities and with high Galactic extinction (see Fig.~\ref{mapcontplot}),
we need to find a subsample where the Galactic extinction and stellar
densities are not correlated. Based on Fig.~\ref{colstdens}, which shows the
extinction values as a function of stellar density for our sample (top
panel), we have chosen the range of $0\fm05 < \ak < 0\fm2$ (shaded area in
the plot) where there is no obvious dependence on stellar densities.
For this subsample of 1395 galaxies we find a dependence of the colours
\jko\ and \hko\ on stellar density (see middle and bottom panel,
respectively). Linear fits to both relationships give
\[(J-K_{\rm s})^{\rm o} = (-0.078\pm0.009) \cdot \log N_*/{\rm deg}^2 + (1.258\pm0.036) \] 
and
\[(H-K_{\rm s})^{\rm o} = (-0.043\pm0.006) \cdot \log N_*/{\rm deg}^2 + (0.449\pm0.024) .\] 
Incidentally, the slopes in these relationships vary only marginally if we
chose a different extinction range for the subsample (\eg increasing the
upper limit up to $\ak = 0\fm5$). Applying these mean relationships as a
correction to the original sample (and using $\log N_*/{\rm deg}^2 = 3.5$
as reference point) we obtain colours independent of stellar density, that is,
$<(J-K_{\rm s})^{\rm o}> = 0\fm962 \pm 0\fm002$ and $<(H-K_{\rm s})^{\rm
  o}> = 0\fm293 \pm 0\fm001$ as well as $<(J-K_{\rm s})^{\rm o,c}> =
1\fm012 \pm 0\fm002$ and $<(H-K_{\rm s})^{\rm o,c}> = 0\fm310 \pm 0\fm001$
(listed as the combined and corrected sample in
Table~\ref{tab2mhistb}). The latter two are in excellent agreement with the
high-latitude sample. The shapes of the distributions of the
fully-corrected colours agree as well (though this is not shown in
Fig.~\ref{2mhist2} to avoid confusion). We conclude that at high stellar
densities, where the likelihood of unsubtracted or undetected faint stars
affecting the galaxy photometry is higher, galaxy colours are too blue and
thus not reliable. For example, at the highest stellar densities where we
still detect galaxies, $\log N_*/{\rm deg}^2 = 4.5$, the \jko\ and
\hko\ colours are too blue by $0\fm08$ and $0\fm04$, respectively.
 
Since galaxy colours are sensitive to foreground extinction, they in turn
can be used to investigate the foreground extinction in more detail
\citep{schroeder07}. We will revisit this in a forthcoming paper (Paper
III).

\section{Summary and discussion} \label{summary} % 6

In this first of a series of papers, we presented a magnitude-limited
catalogue of 2MASX galaxies at low Galactic latitudes and at high
foreground extinction levels. The catalogue supplements the high-latitude
2MRS \citep{huchra12} and 2MTF (\eg \citealt{masters14}) surveys with the
aim to obtain a homogeneous, complete and truly `whole-sky' survey for
cosmic flow analyses (the forthcoming papers will address these aims). We
have used an extinction-corrected magnitude limit of \Ko $= 11\fm25$ which
is brighter than the 2MRS limit of \Ko\ $= 11\fm75$, but the higher
extinctions in our sample would make the parameters of fainter galaxies
less reliable. At all Galactic longitudes, we extracted 2MASX objects with
latitudes $|b|<10\degr$ (the `ZOA sample') and at higher latitudes we
selected all objects with \ebv $>0\fm950$ (the `EBV sample'). All objects
were visually inspected across a wide wavelength range to exclude all non
galaxies from the sample. This results in 3675 and 88 galaxies in the ZOA
and EBV galaxy sample, respectively, with a rejection rate for non-galaxies
in the 2MASX catalogue of 47\% and 82\%, respectively. The completeness of
our catalogue is mainly affected by the stellar density (the completeness
limit lies at $\log N_*/{\rm deg}^2 = 4.5$), while the extinction seems to
have only a small effect, if any, up to $\ak \simeq 2\fm0$. Thus, the NIR
ZoA for bright galaxies covers only 2.4\% of the full sky.  For each galaxy
we give an estimate of its morphological type, based on the likelihood of
it having a disc. Furthermore, we identified which galaxies have optical or
\HI\ radial velocities measurements.

Since the uncertainties in galaxy parameters increase towards the Galactic
plane, we discuss the effects of foreground extinction and high stellar
densities. In particular, the 2MASX photometry in the ZoA is affected by
the high stellar densities which (i) increase the confusion noise and
therewith decrease the sensitivity which can bias the background
subtraction \citep{Jarrett+2000b}, (ii) can cause the centre of the
extraction aperture to be located on a superimposed star instead of the
galaxy, and (iii) sometimes affects the size, ellipticity and position
angle of the extraction aperture. Based on investigations by
\citet{said16b}, \citet{andreon02} and \citet{Kirby+2008}, we conclude that
2MASX magnitudes in the ZoA are on average too bright by at least $0\fm35$
compared to 2MASX galaxies at higher latitudes, an effect which needs to be
taken into account when combining the 2MRS and 2MTF surveys with our
catalogue. The colours as well are sensitive to unresolved stars in the
field, and we show that at the high stellar density level of $\log N_*/{\rm
  deg}^2 = 4.5$ the colours \jko\ and \hko are too blue on average by
$0\fm08$ and $0\fm04$, respectively. We estimate the uncertainty in the
limiting magnitude of $11\fm25$ of our catalogue to be $\sim\!0\fm15$.

Considering the problems of obtaining accurate photometry in star-crowded
areas as well as the underestimation of the 2MASX fluxes due to the short
exposure time, we strongly argue for deeper all-sky surveys with higher
spatial resolution in the NIR. These will be available for the southern
hemisphere once the VHS \citep{mcmahon13} and VVV \citep{minniti10} surveys
are fully completed, but not in the north where the UKIRT Hemisphere Survey
\citep{dye18} does not cover the North Celestial Cap.

Different kinds of Galactic extinction corrections are suggested in the
literature: values extracted from DIRBE/IRAS maps by SFD98 were found to be
too high and need to be corrected (\eg \citealt{schroeder07,schlafly11}).
In addition, the measured size of an obscured galaxy is also affected by
foreground extinction, and an additional correction for the isophotal
diameter is necessary
(\citealt{Cameron1990}, \citealt{riad2010}). While the sample selection of
our main catalogue is based on the SFD98 correction, so as to be as close
as possible to the selection criteria of the 2MRS survey, we also include
the optimal extinction correction to the photometry parameters in our
catalogue and discussion. We also investigated how the sample selection
would change if this optimal extinction correction were applied: Firstly,
because the diameter-dependent extinction correction makes a galaxy
brighter, we additionally extracted all the 2MASX objects which are
brighter than the fully corrected magnitude limit of \Kd $=11\fm25$, but
are fainter than our original limit of \Ko $=11\fm25$. This supplementary
sample comprises 70 and 1 additional galaxies in the ZOA and EBV samples,
respectively. Secondly, 29 and 0 galaxies, respectively, are removed from
the catalogue due to the SF11 correction factor to the extinction
values. While these two effects almost cancel each other out and the
so-called supplementary sample is small, it is nonetheless an important
correction to the main catalogue and should be taken into account in any flow
field analyses to avoid biases due to the patchy distribution of the
supplementary galaxies (see Figure~\ref{mapallplot} in the appendix).

\begin{figure} % 15
\centering
\includegraphics[width=0.45\textwidth]{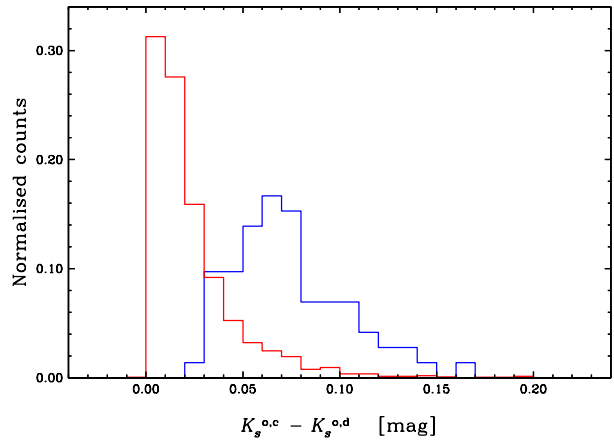}
\caption{Histograms of the difference between \K -band magnitudes after 
  basic extinction correction using the SF11 correction factor $f=0.86$,
  \koc , and magnitudes with the additional diameter correction, \kd , for
  the ZOA sample (red) and EBV sample (blue); only galaxies that are also
  in the 2MRS catalogue are shown.
}
\label{histkd}
\end{figure}

For a truly unbiased analysis, the 2MRS and 2MTF photometry needs to have
the same correction applied. To emphasise the importance of this correction
we show in Fig.~\ref{histkd} the histograms of the {\it additional}
correction due to the diameter-dependent extinction correction for all 2MRS
galaxies in our two samples. Note that for an unbiased comparison, we
applied the SF11 correction factor, $f=0.86$, to the \ko\ values (now
denoted as \koc ).

For the ZOA sample the median difference in magnitude is $0\fm02$, where 34
galaxies (1.4\% out of 2360) deviate more than $0\fm10$. For the EBV
sample, that is, in the overlap region of \ebv\, $= 0\fm950 - 1\fm000$, we
have a median difference of $0\fm07$, and 14 out of the 72 galaxies (19\%)
deviate more than $0\fm10$. Though the numbers and areas affected by
extinctions above $0\fm95$ are small, the omission of this correction even
even a much lower level of, for example, \ebv $=0\fm44$ (or $\ak =0\fm16$)
would imply a {\it systematic} error in magnitude of the order of the
typical uncertainty in galaxy magnitudes, that is, $0\fm20$. Such a
  magnitude difference would result in distances 10\% too large, and
  could affect peculiar velocities by as much as a factor of 2. And since the
foreground extinction is distributed unevenly across the sky this bias
could mimic a non-existent flow field (\eg \citealt{Kolatt1995}). 

We also find that the value of the foreground extinction for a given galaxy
can be uncertain, especially in high extinction areas of molecular clouds,
as it is based on the relatively coarse $6\arcmin$-resolution of the
DIRBE/IRAS maps.  In addition, at $|b|<5\degr$ no FIR sources were
subtracted from these maps due to confusion problems. In this region, we
find three bright FIR sources that clearly bias the local extinction
estimates. We estimate that a further $\approxgt \,22$\% of the galaxies
are likely affected by weaker FIR sources. This effect needs to be taken
into account in cosmic flow analyses since it implies that
extinction-corrected magnitudes of affected galaxies will be systemically
too bright, albeit by a small amount. We have flagged both effects in the
catalogue.

To ensure that a combination of our catalogue with the 2MRS and 2MTF
surveys will not introduce a bias, we compare the extinction-corrected
2MASX diameter and \K -band magnitude of our galaxies with a high-latitude
2MASX galaxy sample with the same magnitude limit. While the median values
of the \K -band magnitude agree well within the errors, the diameters in
the ZoA are systematically too small by $\sim4''$ even when we apply the
diameter-dependent extinction correction. We assume this is likely due to
an incomplete diameter-dependent extinction correction which was derived
from a small sample \citep{riad2010}. Since no such effect for the corrected
magnitudes could be found, we are confident that the sample selection,
which is based on magnitudes, is not affected.
 
Finally, we find that although the galaxy extraction from the 2MASX
catalogue presented here is complete to a given limit in magnitude, the
catalogue itself is missing galaxies in the Galactic bulge area due to the
difficulties of the 2MASX automated galaxy-star separation algorithms to
identify galaxies in regions of high stellar density. NIR surveys with a
better spatial resolution can help reduce the remaining ZoA
(\eg Vista-VVV, \citealt{minniti10}, and UKIDSS-GPS, \citealt{lucas08}),
in particular in conjunction with improved galaxy detection algorithms
  (based on machine learning, \eg \citealt{gonzalez18}).

To further improve the sample we propose to use a semi-automated script to
extract better (2MASS) galaxy photometry (\eg \citealt{said17}) and thus
remove some of the uncertainties in the magnitude limit of the sample (or
sample definition). Furthermore, we find that galaxy photometry in the ZoA
not only requires a correction for Galactic extinction but also that the
increased confusion noise (due to unresolved faint stars) affects the
accuracy of the background subtraction. For example, at high latitudes the
20 mag arcsec$^{-2}$ isophotal \K -band magnitude corresponds roughly to
only $1\sigma$ of the typical background noise in the \K -band
\citep{Jarrett+2000b}, but at low latitudes the increased noise level means
that the isophotal aperture is less well determined. Thus, all photometry
used in cosmic flow field analyses should preferentially be derived from
NIR observations deeper than 2MASX (\eg UKIDSS or VISTA survey data). 

To conclude, a recipe for combining high-latitude surveys like 2MRS and
2MTF with a ZoA survey like ours needs to address the following issues: (a)
full extinction corrections need to be applied in the same manner to all
galaxies, even at moderate extinction levels; (b) possible offsets in the
isophotal magnitudes due to increased background noise in the Galactic
plane need to be corrected for; (c) the variance in photometric quality and
thus the increased photometric uncertainties in localised areas on the sky
needs to be understood and taken into account as a possible bias; (d)
over-estimation of the extinction correction for a significant number of
galaxies due to weak and unresolved FIR sources in the IRAS/DIRBE maps at
low latitudes need to be assessed and taken into account.

In the forthcoming papers we will present \HI\ observations for those
galaxies that do not yet have redshifts and of those with redshifts that
are eligible for the TF analysis. Preparatory work on the application of
the TF relation in the ZoA has been done by \citet{said15}, who determined
that isophotal apertures are less affected by high stellar densities than
total magnitudes, and that axial ratios need to be corrected for
extinction. The final paper in this series will deal with the application
the TF relation to the whole-sky sample (see \citealt{said17} for a pilot
study). Fine-tuning of the Galactic foreground extinction determination
will be discussed in another paper in this series (Schr\"oder \etal , in
prep.).

\section*{Acknowledgments}

The authors thank Khaled Said for the use of his 2MASX\,--\,deep NIR
comparison data and Lucas Macri for access to the most updated 2MRS data
set. The \nan\ Radio Telescope is operated as part of the Paris
Observatory, in association with the Centre National de la Recherche
Scientifique (CNRS) and partially supported by the R\'egion Centre in
France. This publication makes use of data products from the Two Micron All
Sky Survey, which is a joint project of the University of Massachusetts and
the Infrared Processing and Analysis Center, funded by the National
Aeronautics and Space Administration and the National Science
Foundation. This research also has made use of the HyperLeda database, the
NASA/IPAC Extragalactic Database (NED) which is operated by the Jet
Propulsion Laboratory, California Institute of Technology, under contract
with the National Aeronautics and Space Administration and the Sloan
Digital Sky Survey which is managed by the Astrophysical Research
Consortium for the Participating Institutions. ACS and RKK thank the South
African NRF for their financial support.

\bibliographystyle{mn2e} % mn2e.bst for references
\bibliography{ZoA_bibfile_311018} % your file with extension .bib containing references

% the Appendix should come directly after the References

%-----------------------------APPENDIX--------------------------------------------------------

\appendix

\section{Extended catalogue according to optimised extinction corrections }   \label{suppl}

In Sec~\ref{diamcorr} we discuss an additional Galactic foreground
extinction correction for isophotal radii and thus isophotal
magnitudes. The main catalogue includes photometry parameters based on this
additional correction as well as a recommended correction to the
\ebv\ values from the IRAS/DIRBE maps (SF11; Sec.~\ref{badext}). Since the
magnitudes are affected, the sample selection criterion regarding a
limiting magnitude is not ful-filled anymore. Hence, we investigated how
the sample selection would change if this optimal extinction correction
were applied and thus compiled an `extended' catalogue.

Any extinction correction makes galaxy magnitudes brighter. If we apply a
diameter-dependent extinction correction in addition to the conventional
Galactic foreground extinction correction, the magnitude limit of our
sample, \Ko\ $= 11\fm25$ is not longer valid and the sample becomes
incomplete. We therefore need to supplement our catalogue by adding all
those galaxies with \Ko $>11\fm25$ but \Kd $\le 11\fm25$.  The
supplementary sample which contains these galaxies is presented in
Table~\ref{zoaihabtabex} (available online) with the same columns as for
the main catalogue (see Sec.~\ref{results}). It comprises 276 and 14
additional objects for the ZOA and EBV samples, respectively. Applying the
object flags we devised for classifying sources as galaxies
(Table~\ref{zoatabex}, Col.~4) results in 70 and 1 additional galaxies,
respectively.

On the other hand, galaxies will be excluded from the main catalogue since
in the derivation of the optimally extinction-corrected \Kd\ values we
applied the correction factor 0.86 from SF11 to the \ebv\ values used,
which makes magnitudes fainter. Objects thus excluded from the main
catalogue are flagged (Table~\ref{zoatabex}, Col.~7d). For ease of use, we
repeat those lines in Table~\ref{zoaexcltab} (dubbed the exclusion
sample). There are 31 and 1 objects affected in the ZOA and EBV samples,
respectively, of which 29 and 0 are galaxies. In total there are 3716 and
89 galaxies in the `extended' ZOA and EBV samples, respectively, using
\Kd\ $= 11\fm25$ (as opposed to the original 3675 and 88, respectively).

Compared to the main catalogue, the supplementary sample has considerably
fewer matches in the 2MRS catalogue: only 4 objects (1\%) have a 2MRS
counterpart. This is expected as the former includes objects fainter than
the 2MRS cut-off at \K $\le 11\fm75$. Another characterisation of a sample
with fainter objects is the greater uncertainty in galaxy classification
(in particular in our classes 4, 5 and 6), and disc types (uncertain class,
D3, and unknown, `n').  This is in fact one of the reasons why we prefer a
higher magnitude limit for our sample, that is, $11\fm25$ versus $11\fm75$
as used by the 2MRS project. Finally, in the supplementary sample only 6\%
galaxies have published redshifts (versus 70\% in the main catalogue).

Figure~\ref{mapallplot} shows the distribution on the sky of the
supplementary galaxies (magenta circles), which is patchy and more
concentrated towards the Galactic plane compared to the main catalogue
galaxies (grey dots). Galaxies excluded from the main catalogue are shown in
green.

\onecolumn
{\tiny
\input{2M_ihab_comb_ex.tex}
}
%\twocolumn
%
%\onecolumn
{\tiny
\input{2M_comb_excl.tex}
}
\twocolumn

%\addtocounter{table}{+2}
\begin{table}  % A3
\centering
\begin{minipage}{140mm}
\caption{{Flags for identifying galaxies in the supplementary and exclusion samples} \label{tabihabflags}}
\begin{tabular}{llrrrr}
\hline
Flag & Galaxy class & \multicolumn{2}{c}{ZOA} & \multicolumn{2}{c}{EBV} \\
     &              & suppl & excl & suppl & excl \\
\hline
1+2 & definitely         &   46 & 29 &   0 &  0 \\
3 & probably             &   13 &  0 &   1 &  0 \\
4 & possibly             &    9 &  0 &   0 &  0 \\
5 & unknown              &    2 &  0 &   0 &  0 \\
6 & low likelihood       &    5 &  0 &   0 &  0 \\
7 & unlikely             &   27 &  1 &   2 &  1 \\
8+9 & no                 &  145 &  1 &  11 &  0 \\
\hline                                     
total &                & +276 & $-31$ & +14 & $-1$ \\
\hline
\end{tabular}
\end{minipage}
\end{table}

\begin{table}  % A4
\centering
\begin{minipage}{140mm}
\caption{{Disc classification of galaxies in the supplementary and exclusion samples } \label{tabihabdisc}}
\begin{tabular}{llrrrr}
\hline
Class & meaning     & \multicolumn{2}{c}{ZOA} & \multicolumn{2}{c}{EBV} \\
     &              & suppl & excl & suppl & excl \\
\hline
D1  & obvious disc   &   31 & 14 &   1 & 0 \\  %class=1
D2  & disc           &   13 &  4 &   0 & 0 \\  %class=2
D3  & possible disc  &   11 &  7 &   0 & 0 \\  %class=3
D4  & no disc        &    0 &  4 &   0 & 0 \\  %class=4,5
n   & unable to tell &   15 &  0 &   0 & 0 \\  %class=9
\hline                                   
total &                & +70 & $-29$ & +1 & $-0$ \\
\hline
\end{tabular}
\end{minipage}
\end{table}

\begin{figure*} % A1
\centering
\includegraphics[height=0.85\textwidth,angle=270,origin=c]{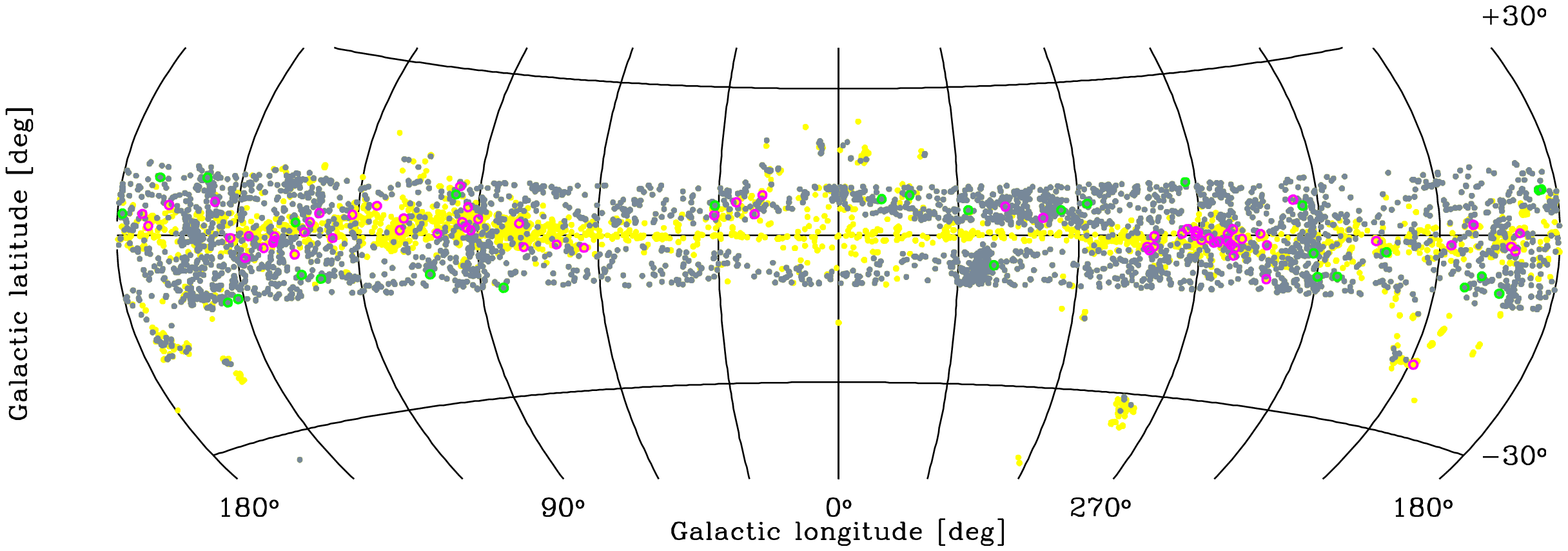}
\vspace{-3.5cm}
\caption{Aitoff projection of the distribution on the sky of the objects in
  the main, supplementary and exclusion samples, in Galactic
  coordinates. Yellow dots represent non-galaxian objects, grey dots
  galaxies from the main catalogue, magenta circles supplementary sample
  galaxies and green circles exclusion sample galaxies.
}
\label{mapallplot}
\end{figure*}

\onecolumn

% -------------------------------- Tables ---------------------------------------------------
%

%\newpage
%\clearpage 
%
%\section{Online material }   \label{online}
%
%\onecolumn
%{\tiny
%%\input{2M_comb.tex}
%}
%
%{\tiny
%%\input{2M_ihab_comb.tex}
%}
%%\newpage
%%\clearpage 

\bsp

\label{lastpage}

\end{document}

%% file: 2M_comb_ex.tex
\begin{landscape}
%\begin{longtable}{lrrrcclccccccllcclrrrrrrrrrrrrrrr}
\begin{longtable}{@{\extracolsep{1.0mm}}l@{\extracolsep{1.0mm}}r@{\extracolsep{1.0mm}}r@{\extracolsep{1.0mm}}r@{\extracolsep{1.0mm}}c@{\extracolsep{1.0mm}}c@{\extracolsep{1.0mm}}l@{\extracolsep{1.0mm}}c@{\extracolsep{1.0mm}}c@{\extracolsep{1.0mm}}c@{\extracolsep{1.0mm}}c@{\extracolsep{1.0mm}}c@{\extracolsep{1.0mm}}c@{\extracolsep{1.0mm}}l@{\extracolsep{1.0mm}}l@{\extracolsep{0.0mm}}c@{\extracolsep{1.0mm}}c@{\extracolsep{1.0mm}}l@{\extracolsep{1.0mm}}r@{\extracolsep{2.0mm}}r@{\extracolsep{2.0mm}}r@{\extracolsep{1.0mm}}r@{\extracolsep{1.0mm}}r@{\extracolsep{2.0mm}}r@{\extracolsep{2.0mm}}r@{\extracolsep{2.0mm}}r@{\extracolsep{1.0mm}}r@{\extracolsep{1.0mm}}r@{\extracolsep{1.0mm}}r@{\extracolsep{1.0mm}}r@{\extracolsep{1.0mm}}r@{\extracolsep{1.0mm}}r@{\extracolsep{1.0mm}}r}
\caption{{\normalsize 2MASS ZoA sample; example page, the full table is
    available online} \label{zoatabex}}\\
\hline
\noalign{\smallskip}
2MASX J          &  $l$ &  $b$ & EBV & \multicolumn{3}{c}{Object} & \multicolumn{4}{c}{Sample} & Disc  & 2MRS & \multicolumn{2}{c}{Velocity} & NIR  & Photom & Ext  & $K_{20}$ & $H$-$K$ & $J$-$K$ & vc  & $a$ & $b/a$ & st.d. & \ko & ($H$-$K)^o$ & ($J$-$K)^o$  &$\ak$& $a^{\rm d}$& \kd &($H$-$K)^{o,c}$ & ($J$-$K)^{o,c}$ \\ 
                 &  deg &  deg & mag & class & off & flg          &  gal & TF   & obs  & ext'd & type  &      & Opt    & HI                  & flg  & flg    & flg  & mag      & mag     &  mag    &     & ''  &       &       & mag & mag &  mag & mag & $''$       & mag & mag &  mag \\
(1)              & (2a) & (2b) & (3) & (4)   & (5) & (6)          & (7a) & (7b) & (7c) & (7d)  & (8)   & (9)  & (10a)  & (10b)               & (11) & (12)   & (13) & (14)     & (15)    & (16)    &(17) & (18)& (19)  & (20)  & (21)& (22)& (23) & (24)& (25)       & (26)& (27) & (28) \\
\noalign{\smallskip}
\hline
\noalign{\smallskip}
\endfirsthead
\caption{continued.}\\
\hline
\noalign{\smallskip}
2MASX J          &  $l$ &  $b$ & EBV & \multicolumn{3}{c}{Object} & \multicolumn{4}{c}{Sample} & Disc  & 2MRS & \multicolumn{2}{c}{Velocity} & NIR  & Photom & Ext  & $K_{20}$ & $H$-$K$ & $J$-$K$ & vc  & $a$ & $b/a$ & st.d. & \ko & ($H$-$K)^o$ & ($J$-$K)^o$  &$\ak$& $a^{\rm d}$& \kd &($H$-$K)^{o,c}$ & ($J$-$K)^{o,c}$ \\ 
                 &  deg &  deg & mag & class & off & flg          &  gal & TF   & obs  & ext'd & type  &      & Opt    & HI                  & flg  & flg    & flg  & mag      & mag     &  mag    &     & ''  &       &       & mag & mag &  mag & mag & $''$       & mag & mag &  mag \\
(1)              & (2a) & (2b) & (3) & (4)   & (5) & (6)          & (7a) & (7b) & (7c) & (7d)  & (8)   & (9)  & (10a)  & (10b)               & (11) & (12)   & (13) & (14)     & (15)    & (16)    &(17) & (18)& (19)  & (20)  & (21)& (22)& (23) & (24)& (25)       & (26)& (27) & (28) \\
\noalign{\smallskip}
\hline
\noalign{\smallskip}
\endhead
\noalign{\smallskip}
\hline
\endfoot
%\noalign{\smallskip}
%ID                &    l     &$   b    $ &$   E(B-V)$  & xv & o &  ps & gal & tf &obs &ihb &  cl &tmrs & vo  & vh   &  U &   ph &  ext &$    K20    $&$  H-K   $&$  J-K   $&$ vc $&       a &$sup_ba  $&$stardens $&$   ko_as  $&$ hko_as $&$ jko_as $&  ak_sf  &$   a2d_sf $&$   kd_sf  $&$ hko_sf $&$  jko_sf $ \\   
00000637$+$5319136 &  115.217 &$  -8.780$ &$    0.300$  &  1 &$-$& $-$ &  g  &$-$ &$-$ &$-$ &  D4 &  c  & o   & $- $ & $-$&  $-$ & $-$  &$   11.196  $&$  0.42  $&$  1.15  $&$  1 $&    35.0 &$  0.90  $&$   3.59  $&$   11.084 $&$   0.37 $&$   1.00 $&   0.09  &$     35.6 $&$   11.09  $&$   0.38 $&$   1.02  $ \\   
00000702$+$6717254 &  117.996 &$   4.912$ &$   10.331$  &  9 &$-$& $-$ & $-$ &$-$ &$-$ &$-$ & $-$ & $-$ &$- $ & $- $ & $-$&  $-$ & $-$  &$   13.614  $&$  0.53  $&$  1.38  $&$ -1 $&    11.0 &$  1.00  $&$   3.63  $&$    9.765 $&$  -1.18 $&$  -3.75 $&   3.20  &$   \cdots $&$  \cdots  $&$  -0.98 $&$  -3.07  $ \\   
00001859$+$6716444 &  118.012 &$   4.897$ &$    9.743$  &  9 &$-$& $-$ & $-$ &$-$ &$-$ &$-$ & $-$ & $-$ &$- $ & $- $ & $-$&  $-$ & $-$  &$   14.306  $&$  0.36  $&$  1.30  $&$ -1 $&    10.0 &$  1.00  $&$   3.63  $&$   10.676 $&$  -1.26 $&$  -3.54 $&   3.02  &$   \cdots $&$  \cdots  $&$  -1.06 $&$  -2.89  $ \\   
00002049$+$6717044 &  118.016 &$   4.902$ &$    9.880$  &  9 &$-$& $-$ & $-$ &$-$ &$-$ &$-$ & $-$ & $-$ &$- $ & $- $ & $-$&  $-$ & $-$  &$   13.371  $&$  0.60  $&$  1.61  $&$ -1 $&    19.4 &$  0.54  $&$   3.63  $&$    9.690 $&$  -1.04 $&$  -3.30 $&   3.06  &$   \cdots $&$  \cdots  $&$  -0.84 $&$  -2.65  $ \\   
00003760$+$6735164 &  118.103 &$   5.194$ &$    4.645$  &  8 &$-$& $-$ & $-$ &$-$ &$-$ &$-$ & $-$ & $-$ &$- $ & $- $ & $-$&  $-$ & $-$  &$   12.242  $&$  0.50  $&$  1.47  $&$ -2 $&    22.6 &$  0.34  $&$   3.91  $&$   10.511 $&$  -0.27 $&$  -0.83 $&   1.44  &$   \cdots $&$  \cdots  $&$  -0.18 $&$  -0.53  $ \\   
00003769$+$6710184 &  118.021 &$   4.786$ &$    6.767$  &  8 &$-$& $-$ & $-$ &$-$ &$-$ &$-$ & $-$ & $-$ &$- $ & $- $ & $-$&  $-$ & $-$  &$   12.635  $&$  1.11  $&$ \cdots $&$  1 $&    37.2 &$  0.46  $&$   3.63  $&$   10.114 $&$  -0.01 $&$ \cdots $&   2.10  &$   \cdots $&$  \cdots  $&$   0.12 $&$ \cdots  $ \\   
00004194$+$6838344 &  118.318 &$   6.227$ &$    0.957$  &  7 &$-$& $-$ & $-$ &$-$ &$-$ &$-$ & $-$ & $-$ &$- $ & $- $ & $-$&  $-$ & $-$  &$    9.622  $&$  1.60  $&$  3.34  $&$  1 $&    23.2 &$  0.78  $&$   3.67  $&$    9.265 $&$   1.44 $&$   2.87 $&   0.30  &$   \cdots $&$  \cdots  $&$   1.46 $&$   2.93  $ \\   
00004647$+$6732523 &  118.109 &$   5.152$ &$    6.505$  &  8 &$-$& $-$ & $-$ &$-$ &$-$ &$-$ & $-$ & $-$ &$- $ & $- $ & $-$&  $-$ & $-$  &$   10.986  $&$  1.17  $&$  3.21  $&$  1 $&    42.4 &$  0.62  $&$   3.91  $&$    8.562 $&$   0.09 $&$  -0.02 $&   2.01  &$   \cdots $&$  \cdots  $&$   0.22 $&$   0.41  $ \\   
00013723$+$6436560 &  117.623 &$   2.259$ &$    4.572$  &  9 &$-$& $-$ & $-$ &$-$ &$-$ &$-$ & $-$ & $-$ &$- $ & $- $ & $-$&  $-$ & $-$  &$   12.737  $&$  0.35  $&$  1.33  $&$ -2 $&    17.8 &$  0.70  $&$   3.94  $&$   11.033 $&$  -0.41 $&$  -0.95 $&   1.42  &$   \cdots $&$  \cdots  $&$  -0.32 $&$  -0.64  $ \\   
00013755$+$6437360 &  117.626 &$   2.270$ &$    4.633$  &  9 &$-$& $-$ & $-$ &$-$ &$-$ &$-$ & $-$ & $-$ &$- $ & $- $ & $-$&  $-$ & $-$  &$   11.735  $&$  0.41  $&$  1.23  $&$  1 $&    37.2 &$  0.80  $&$   3.94  $&$   10.009 $&$  -0.36 $&$  -1.08 $&   1.44  &$   \cdots $&$  \cdots  $&$  -0.27 $&$  -0.77  $ \\   
00014201$+$6724323 &  118.169 &$   4.999$ &$    7.957$  &  8 &$-$& $-$ & $-$ &$-$ &$-$ &$-$ & $-$ & $-$ &$- $ & $- $ & $-$&  $-$ & $-$  &$   13.778  $&$  0.56  $&$  1.28  $&$ -1 $&    13.6 &$  1.00  $&$   3.94  $&$   10.813 $&$  -0.76 $&$  -2.67 $&   2.46  &$   \cdots $&$  \cdots  $&$  -0.60 $&$  -2.14  $ \\   
00014668$+$6708413 &  118.126 &$   4.738$ &$    4.218$  &  8 &$-$& $-$ & $-$ &$-$ &$-$ &$-$ & $-$ & $-$ &$- $ & $- $ & $-$&  $-$ & $-$  &$   11.590  $&$  1.18  $&$  2.17  $&$  1 $&    61.8 &$  0.52  $&$   3.79  $&$   10.018 $&$   0.48 $&$   0.07 $&   1.31  &$   \cdots $&$  \cdots  $&$   0.57 $&$   0.35  $ \\   
00031331$+$5352149 &  115.783 &$  -8.330$ &$    0.299$  &  1 &$-$& $-$ &  g  &$-$ &$-$ &$-$ &  D3 &  c  & o   & $- $ & $-$&  $-$ & $-$  &$   10.252  $&$  0.37  $&$  1.20  $&$  1 $&    48.6 &$  0.78  $&$   3.57  $&$   10.141 $&$   0.32 $&$   1.05 $&   0.09  &$     49.3 $&$   10.15  $&$   0.33 $&$   1.07  $ \\   
00033726$+$6919064 &  118.707 &$   6.840$ &$    0.808$  &  2 &$-$& $-$ &  g  &$-$ &$-$ &$-$ &  D4 &  c  & o   & $- $ & $-$&  $-$ & $-$  &$   11.480  $&$  0.40  $&$  1.50  $&$  1 $&    26.4 &$  0.70  $&$   3.60  $&$   11.179 $&$   0.27 $&$   1.10 $&   0.25  &$     28.1 $&$   11.19  $&$   0.28 $&$   1.15  $ \\   
00034142$+$7036434 &  118.955 &$   8.110$ &$    0.789$  &  1 &$-$& $-$ &  g  &$-$ &$-$ &$-$ &  D1 &  c  & o   & $- $ & $-$&  $-$ & $-$  &$   11.544  $&$  0.35  $&$  1.29  $&$  1 $&    36.6 &$  0.68  $&$   3.55  $&$   11.250 $&$   0.22 $&$   0.90 $&   0.24  &$     39.4 $&$   11.24  $&$   0.23 $&$   0.95  $ \\   
\end{longtable}
\end{landscape}

%% file: 2M_ihab_comb_ex.tex
\begin{landscape}
%\begin{longtable}{lrrrcclccccccllcclrrrrrrrrrrrrrrr}
\begin{longtable}{@{\extracolsep{1.0mm}}l@{\extracolsep{1.0mm}}r@{\extracolsep{1.0mm}}r@{\extracolsep{1.0mm}}r@{\extracolsep{1.0mm}}c@{\extracolsep{1.0mm}}c@{\extracolsep{1.0mm}}l@{\extracolsep{1.0mm}}c@{\extracolsep{1.0mm}}c@{\extracolsep{1.0mm}}c@{\extracolsep{1.0mm}}c@{\extracolsep{1.0mm}}c@{\extracolsep{1.0mm}}c@{\extracolsep{1.0mm}}l@{\extracolsep{1.0mm}}l@{\extracolsep{0.0mm}}c@{\extracolsep{1.0mm}}c@{\extracolsep{1.0mm}}l@{\extracolsep{1.0mm}}r@{\extracolsep{2.0mm}}r@{\extracolsep{2.0mm}}r@{\extracolsep{1.0mm}}r@{\extracolsep{1.0mm}}r@{\extracolsep{2.0mm}}r@{\extracolsep{2.0mm}}r@{\extracolsep{2.0mm}}r@{\extracolsep{1.0mm}}r@{\extracolsep{1.0mm}}r@{\extracolsep{1.0mm}}r@{\extracolsep{1.0mm}}r@{\extracolsep{1.0mm}}r@{\extracolsep{1.0mm}}r@{\extracolsep{1.0mm}}r}
\caption{{\normalsize 2MASS ZoA supplementary sample; example page, the full table is
    available online} \label{zoaihabtabex}}\\
\hline
\noalign{\smallskip}
2MASX J          &  $l$ &  $b$ & EBV & \multicolumn{3}{c}{Object} & \multicolumn{4}{c}{Sample} & Disc  & 2MRS & \multicolumn{2}{c}{Velocity} & NIR  & Photom & Ext  & $K_{20}$ & $H$-$K$ & $J$-$K$ & vc  & $a$ & $b/a$ & st.d. & \ko & ($H$-$K)^o$ & ($J$-$K)^o$  &$\ak$& $a^{\rm d}$& \kd &($H$-$K)^{o,c}$ & ($J$-$K)^{o,c}$ \\ 
                 &  deg &  deg & mag & class & off & flg          &  gal & TF   & obs  & ext'd & type  &      & Opt    & HI                  & flg  & flg    & flg  & mag      & mag     &  mag    &     & ''  &       &       & mag & mag &  mag & mag & $''$       & mag & mag &  mag \\
(1)              & (2a) & (2b) & (3) & (4)   & (5) & (6)          & (7a) & (7b) & (7c) & (7d)  & (8)   & (9)  & (10a)  & (10b)               & (11) & (12)   & (13) & (14)     & (15)    & (16)    &(17) & (18)& (19)  & (20)  & (21)& (22)& (23) & (24)& (25)       & (26)& (27) & (28) \\
\noalign{\smallskip}
\hline
\noalign{\smallskip}
\endfirsthead
\caption{continued.}\\
\hline
\noalign{\smallskip}
2MASX J          &  $l$ &  $b$ & EBV & \multicolumn{3}{c}{Object} & \multicolumn{4}{c}{Sample} & Disc  & 2MRS & \multicolumn{2}{c}{Velocity} & NIR  & Photom & Ext  & $K_{20}$ & $H$-$K$ & $J$-$K$ & vc  & $a$ & $b/a$ & st.d. & \ko & ($H$-$K)^o$ & ($J$-$K)^o$  &$\ak$& $a^{\rm d}$& \kd &($H$-$K)^{o,c}$ & ($J$-$K)^{o,c}$ \\ 
                 &  deg &  deg & mag & class & off & flg          &  gal & TF   & obs  & ext'd & type  &      & Opt    & HI                  & flg  & flg    & flg  & mag      & mag     &  mag    &     & ''  &       &       & mag & mag &  mag & mag & $''$       & mag & mag &  mag \\
(1)              & (2a) & (2b) & (3) & (4)   & (5) & (6)          & (7a) & (7b) & (7c) & (7d)  & (8)   & (9)  & (10a)  & (10b)               & (11) & (12)   & (13) & (14)     & (15)    & (16)    &(17) & (18)& (19)  & (20)  & (21)& (22)& (23) & (24)& (25)       & (26)& (27) & (28) \\
\noalign{\smallskip}
\hline
\noalign{\smallskip}
\endhead
\noalign{\smallskip}
\hline
\endfoot
%\noalign{\smallskip}
%ID                &    l     &$   b    $ &$   E(B-V)$  & xv & o &  ps & gal & tf &obs &ihb &  cl  &tmrs & vo  & vh   &  U &   ph &  ext &$    K20    $&$  H-K   $&$  J-K   $&$ vc $&       a &$sup_ba  $&$stardens $&$   ko_as  $&$ hko_as $&$ jko_as $&  ak_sf  &$   a2d_sf $&$   kd_sf  $&$ hko_sf $&$  jko_sf $ \\   
00012132$+$6706042 &  118.077 &$   4.703$ &$    4.827$  &  6 &$-$& $-$ & $-$ &$-$ &$-$ &$-$ & $- $ & $-$ &$- $ & $- $ & $-$&  $-$ &  e   &$   13.262  $&$ \cdots $&$ \cdots $&$ -1 $&    17.0 &$  1.00  $&$   3.79  $&$   11.463 $&$  \cdots$&$  \cdots$&   1.50  &$   \cdots $&$  \cdots  $&$ \cdots $&$ \cdots  $ \\  
00021688$+$6719539 &  118.209 &$   4.912$ &$    4.737$  &  8 &$-$& $-$ & $-$ &$-$ &$-$ &$-$ & $- $ & $-$ &$- $ & $- $ & $-$&  $-$ & $-$  &$   13.232  $&$  0.57  $&$  1.65  $&$ -1 $&    12.8 &$  1.00  $&$   4.13  $&$   11.467 $&$  -0.21 $&$  -0.71 $&   1.47  &$   \cdots $&$  \cdots  $&$  -0.12 $&$  -0.39  $ \\  
00102080$+$6521098 &  118.662 &$   2.822$ &$    4.784$  &  7 &$-$& $-$ & $-$ &$-$ &$-$ &$-$ & $- $ & $-$ &$- $ & $- $ & $-$&  $-$ & $-$  &$   13.118  $&$  1.10  $&$  1.61  $&$ -1 $&    17.6 &$  1.00  $&$   3.83  $&$   11.335 $&$   0.30 $&$  -0.77 $&   1.48  &$   \cdots $&$  \cdots  $&$   0.40 $&$  -0.45  $ \\  
00235245$+$6610283 &  120.149 &$   3.452$ &$    4.307$  &  7 &$-$& $-$ & $-$ &$-$ &$-$ &$-$ & $- $ & $-$ &$- $ & $- $ & $-$&  $-$ & $-$  &$   13.105  $&$  0.77  $&$ \cdots $&$  1 $&    20.4 &$  0.54  $&$   3.86  $&$   11.500 $&$   0.06 $&$  \cdots$&   1.33  &$   \cdots $&$  \cdots  $&$   0.15 $&$ \cdots  $ \\  
00235531$+$6605253 &  120.145 &$   3.368$ &$    4.505$  &  8 &$-$& $-$ & $-$ &$-$ &$-$ &$-$ & $- $ & $-$ &$- $ & $- $ & $-$&  $-$ & $-$  &$   13.138  $&$  1.11  $&$ \cdots $&$  1 $&    17.0 &$  0.48  $&$   3.86  $&$   11.459 $&$   0.36 $&$  \cdots$&   1.40  &$   \cdots $&$  \cdots  $&$   0.45 $&$ \cdots  $ \\  
00262333$+$6446048 &  120.265 &$   2.026$ &$    4.632$  &  8 &$-$& $-$ & $-$ &$-$ &$-$ &$-$ & $- $ & $-$ &$- $ & $- $ & $-$&  $-$ & $-$  &$   13.138  $&$ \cdots $&$  1.60  $&$ -1 $&    10.0 &$  1.00  $&$   3.90  $&$   11.412 $&$  \cdots$&$  -0.70 $&   1.43  &$   \cdots $&$  \cdots  $&$ \cdots $&$  -0.40  $ \\  
00265579$+$6510278 &  120.361 &$   2.425$ &$    2.480$  &  6 &$-$& $-$ & $-$ &$-$ &$-$ &$-$ & $- $ & $-$ &$- $ & $- $ & $-$&  $-$ & $-$  &$   12.253  $&$  0.74  $&$  1.86  $&$ -2 $&    20.0 &$  0.76  $&$   3.81  $&$   11.329 $&$   0.33 $&$   0.63 $&   0.77  &$   \cdots $&$  \cdots  $&$   0.38 $&$   0.80  $ \\  
00283778$+$6527431 &  120.564 &$   2.695$ &$    2.976$  &  7 &$-$& $-$ & $-$ &$-$ &$-$ &$-$ & $- $ & $-$ &$- $ & $- $ & $-$&  $-$ & $-$  &$   12.547  $&$  1.01  $&$  2.31  $&$  1 $&    23.2 &$  0.50  $&$   3.77  $&$   11.438 $&$   0.52 $&$   0.83 $&   0.92  &$   \cdots $&$  \cdots  $&$   0.58 $&$   1.02  $ \\  
00291882$+$6419524 &  120.538 &$   1.563$ &$    2.850$  &  9 &$-$& $-$ & $-$ &$-$ &$-$ &$-$ & $- $ & $-$ &$- $ & $- $ & $-$&  $-$ & $-$  &$   12.456  $&$  1.12  $&$ \cdots $&$  1 $&    34.8 &$  0.78  $&$   3.87  $&$   11.394 $&$   0.65 $&$  \cdots$&   0.88  &$   \cdots $&$  \cdots  $&$   0.71 $&$ \cdots  $ \\  
00353673$+$6618340 &  121.340 &$   3.487$ &$    1.967$  &  2 &$-$& $-$ &  g  & t  &$-$ &$-$ &  D2  & $-$ &$- $ & $- $ & $-$&  $-$ &  e   &$   11.988  $&$  0.62  $&$  1.77  $&$  1 $&    27.2 &$  0.40  $&$   3.80  $&$   11.255 $&$   0.29 $&$   0.80 $&   0.61  &$     33.8 $&$   11.22  $&$   0.33 $&$   0.93  $ \\  
\end{longtable}
\end{landscape}

%% file: 2M_comb_excl.tex
\begin{landscape}
%\begin{longtable}{lrrrcclccccccllcclrrrrrrrrrrrrrrr}
\begin{longtable}{@{\extracolsep{1.0mm}}l@{\extracolsep{1.0mm}}r@{\extracolsep{1.0mm}}r@{\extracolsep{1.0mm}}r@{\extracolsep{1.0mm}}c@{\extracolsep{1.0mm}}c@{\extracolsep{1.0mm}}l@{\extracolsep{1.0mm}}c@{\extracolsep{1.0mm}}c@{\extracolsep{1.0mm}}c@{\extracolsep{1.0mm}}c@{\extracolsep{1.0mm}}c@{\extracolsep{1.0mm}}c@{\extracolsep{1.0mm}}l@{\extracolsep{1.0mm}}l@{\extracolsep{0.0mm}}c@{\extracolsep{1.0mm}}c@{\extracolsep{1.0mm}}l@{\extracolsep{1.0mm}}r@{\extracolsep{2.0mm}}r@{\extracolsep{2.0mm}}r@{\extracolsep{1.0mm}}r@{\extracolsep{1.0mm}}r@{\extracolsep{2.0mm}}r@{\extracolsep{2.0mm}}r@{\extracolsep{2.0mm}}r@{\extracolsep{1.0mm}}r@{\extracolsep{1.0mm}}r@{\extracolsep{1.0mm}}r@{\extracolsep{1.0mm}}r@{\extracolsep{1.0mm}}r@{\extracolsep{1.0mm}}r@{\extracolsep{1.0mm}}r}
\caption{{\normalsize a) 2MASS exclusion ZoA sample } \label{zoaexcltab}}\\
\hline
\noalign{\smallskip}
2MASX J          &  $l$ &  $b$ & EBV & \multicolumn{3}{c}{Object} & \multicolumn{4}{c}{Sample} & Disc  & 2MRS & \multicolumn{2}{c}{Velocity} & NIR  & Photom & Ext  & $K_{20}$ & $H$-$K$ & $J$-$K$ & vc  & $a$ & $b/a$ & st.d. & \ko & ($H$-$K)^o$ & ($J$-$K)^o$  &$\ak$& $a^{\rm d}$& \kd &($H$-$K)^{o,c}$ & ($J$-$K)^{o,c}$ \\ 
                 &  deg &  deg & mag & class & off & flg          &  gal & TF   & obs  & ext'd & type  &      & Opt    & HI                  & flg  & flg    & flg  & mag      & mag     &  mag    &     & ''  &       &       & mag & mag &  mag & mag & $''$       & mag & mag &  mag \\
(1)              & (2a) & (2b) & (3) & (4)   & (5) & (6)          & (7a) & (7b) & (7c) & (7d)  & (8)   & (9)  & (10a)  & (10b)               & (11) & (12)   & (13) & (14)     & (15)    & (16)    &(17) & (18)& (19)  & (20)  & (21)& (22)& (23) & (24)& (25)       & (26)& (27) & (28) \\
\noalign{\smallskip}
\hline
\noalign{\smallskip}
\endfirsthead
\caption{\normalsize a) continued.}\\
\hline
\noalign{\smallskip}
2MASX J          &  $l$ &  $b$ & EBV & \multicolumn{3}{c}{Object} & \multicolumn{4}{c}{Sample} & Disc  & 2MRS & \multicolumn{2}{c}{Velocity} & NIR  & Photom & Ext  & $K_{20}$ & $H$-$K$ & $J$-$K$ & vc  & $a$ & $b/a$ & st.d. & \ko & ($H$-$K)^o$ & ($J$-$K)^o$  &$\ak$& $a^{\rm d}$& \kd &($H$-$K)^{o,c}$ & ($J$-$K)^{o,c}$ \\ 
                 &  deg &  deg & mag & class & off & flg          &  gal & TF   & obs  & ext'd & type  &      & Opt    & HI                  & flg  & flg    & flg  & mag      & mag     &  mag    &     & ''  &       &       & mag & mag &  mag & mag & $''$       & mag & mag &  mag \\
(1)              & (2a) & (2b) & (3) & (4)   & (5) & (6)          & (7a) & (7b) & (7c) & (7d)  & (8)   & (9)  & (10a)  & (10b)               & (11) & (12)   & (13) & (14)     & (15)    & (16)    &(17) & (18)& (19)  & (20)  & (21)& (22)& (23) & (24)& (25)       & (26)& (27) & (28) \\
\noalign{\smallskip}
\hline
\noalign{\smallskip}
\endhead
\noalign{\smallskip}
\hline
\endfoot
%\noalign{\smallskip}
%ID                &    l     &$   b    $ &$   E(B-V)$  & xv & o &  ps & gal & tf &obs &ihb &  cl &tmrs & vo  & vh   &  U &   ph &  ext &$    K20    $&$  H-K   $&$  J-K   $&$ vc $&       a &$sup_ba  $&$stardens $&$   ko_as  $&$ hko_as $&$ jko_as $&  ak_sf  &$   a2d_sf $&$   kd_sf  $&$ hko_sf $&$  jko_sf $ \\   
01394085$+$5507549 &  129.846 &$  -7.091$ &$    0.319$  &  1 &$-$& $-$ &  g  & t  & N  & *  &  D2 &  c  & o   & $- $ & $-$&  $-$ & $-$  &$   11.368  $&$  0.49  $&$  1.25  $&$  1 $&    33.4 &$  0.46  $&$   3.55  $&$   11.249 $&$   0.44 $&$   1.09 $&   0.10  &$     34.0 $&$   11.26  $&$   0.45 $&$   1.11  $ \\   
02010098$+$6337038 &  130.569 &$   1.783$ &$    1.271$  &  7 &$-$& $-$ & $-$ &$-$ &$-$ & *  & $-$ & $-$ &$- $ & $- $ & $-$&  $-$ & $-$  &$   11.713  $&$  1.02  $&$  2.36  $&$  1 $&    16.4 &$  0.84  $&$   3.82  $&$   11.239 $&$   0.81 $&$   1.73 $&   0.39  &$   \cdots $&$  \cdots  $&$   0.84 $&$   1.81  $ \\   
02122164$+$5441091 &  134.535 &$  -6.352$ &$    0.270$  &  1 &$-$& $-$ &  g  &$-$ &$-$ & *  &  D4 &  c  & o   & $- $ & $-$&  $-$ & $-$  &$   11.350  $&$  0.56  $&$  1.23  $&$  1 $&    28.2 &$  0.78  $&$   3.56  $&$   11.249 $&$   0.51 $&$   1.09 $&   0.08  &$     28.6 $&$   11.26  $&$   0.52 $&$   1.11  $ \\   
02422242$+$6213323 &  135.465 &$   2.084$ &$    0.525$  &  1 &$-$& $-$ &  g  &$-$ & N  & *  &  D3 & $-$ &$- $ & $- $ & $-$&  $-$ & $-$  &$   11.445  $&$  0.49  $&$  1.38  $&$  1 $&    23.6 &$  0.82  $&$   3.79  $&$   11.249 $&$   0.40 $&$   1.12 $&   0.16  &$     24.4 $&$   11.26  $&$   0.41 $&$   1.15  $ \\   
03223065$+$6329185 &  138.871 &$   5.394$ &$    1.654$  &  9 &$-$& $-$ & $-$ &$-$ &$-$ & *  & $-$ & $-$ &$- $ & $- $ & $-$&  $-$ & $-$  &$   11.854  $&$  1.63  $&$  3.21  $&$  1 $&    16.6 &$  0.58  $&$   3.63  $&$   11.238 $&$   1.35 $&$   2.39 $&   0.51  &$   \cdots $&$  \cdots  $&$   1.39 $&$   2.50  $ \\   
03351760$+$4401160 &  151.492 &$  -9.578$ &$    0.313$  &  1 &$-$& $-$ &  g  & t  & N  & *  &  D2 &  c  & o   & $- $ & $-$&  $-$ & $-$  &$   11.364  $&$  0.33  $&$  1.16  $&$  1 $&    34.8 &$  0.44  $&$   3.53  $&$   11.247 $&$   0.28 $&$   1.00 $&   0.10  &$     35.4 $&$   11.26  $&$   0.28 $&$   1.02  $ \\   
03460109$+$4201053 &  154.307 &$  -9.976$ &$    0.317$  &  1 &$-$& $-$ &  g  &$-$ &$-$ & *  &  D2 &  c  & o   & $- $ & $-$&  $-$ & $-$  &$   11.366  $&$  0.31  $&$  1.14  $&$  1 $&    29.0 &$  0.78  $&$   3.46  $&$   11.248 $&$   0.26 $&$   0.99 $&   0.10  &$     29.5 $&$   11.26  $&$   0.26 $&$   1.01  $ \\   
05231302$+$5112435 &  158.869 &$   8.400$ &$    0.389$  &  1 &$-$& $-$ &  g  &$-$ &$-$ & *  &  D1 &  c  & o   & $- $ & $-$&  $-$ & $-$  &$   11.389  $&$  0.31  $&$  1.19  $&$  1 $&    31.0 &$  0.90  $&$   3.52  $&$   11.244 $&$   0.25 $&$   1.00 $&   0.12  &$     31.8 $&$   11.25  $&$   0.26 $&$   1.02  $ \\   
05454092$+$1306184 &  193.613 &$  -8.158$ &$    0.507$  &  1 &$-$& $-$ &  g  & t  &$-$ & *  &  D1 &  c  &$- $ &  h   & $-$&  $-$ & $-$  &$   11.431  $&$  0.34  $&$  1.22  $&$  1 $&    34.0 &$  0.32  $&$   3.43  $&$   11.242 $&$   0.26 $&$   0.97 $&   0.16  &$     35.2 $&$   11.25  $&$   0.27 $&$   1.00  $ \\   
05540495$+$3127485 &  178.768 &$   2.866$ &$    0.492$  &  1 &$-$& $-$ &  g  &$-$ & N  & *  &  D3 & $-$ &$- $ & $- $ &  U &  $-$ & $-$  &$   11.422  $&$  0.44  $&$  1.31  $&$  1 $&    25.0 &$  0.72  $&$   3.68  $&$   11.239 $&$   0.36 $&$   1.07 $&   0.15  &$     25.8 $&$   11.25  $&$   0.37 $&$   1.10  $ \\   
05561062$+$4100185 &  170.662 &$   7.979$ &$    0.344$  &  1 &$-$& $-$ &  g  &$-$ &$-$ & *  &  D1 &  c  & o   & $- $ & $-$&  $-$ & $-$  &$   11.372  $&$  0.41  $&$  1.20  $&$  1 $&    39.2 &$  0.66  $&$   3.44  $&$   11.244 $&$   0.36 $&$   1.03 $&   0.11  &$     40.0 $&$   11.25  $&$   0.36 $&$   1.05  $ \\   
06041759$+$0946030 &  198.807 &$  -5.843$ &$    0.436$  &  1 &$-$& $-$ &  g  &$-$ &$-$ & *  &  D3 &  c  & o   & $- $ & $-$&  $-$ & $-$  &$   11.412  $&$  0.42  $&$  1.30  $&$  1 $&    35.2 &$  0.82  $&$   3.60  $&$   11.250 $&$   0.35 $&$   1.08 $&   0.14  &$     36.3 $&$   11.26  $&$   0.36 $&$   1.11  $ \\   
06051642$+$0535242 &  202.616 &$  -7.631$ &$    0.442$  &  1 &$-$& $-$ &  g  & t  &$-$ & *  &  D1 &  c  & o   & $- $ & $-$&  $-$ & $-$  &$   11.415  $&$  0.38  $&$  1.22  $&$  1 $&    45.4 &$  0.50  $&$   3.52  $&$   11.250 $&$   0.31 $&$   1.00 $&   0.14  &$     46.9 $&$   11.25  $&$   0.32 $&$   1.03  $ \\   
06181338$+$2844455 &  183.687 &$   6.120$ &$    0.566$  &  1 &$-$& $-$ &  g  &$-$ &$-$ & *  &  D3 &  c  & o   & $- $ & $-$&  $-$ & $-$  &$   11.455  $&$  0.42  $&$  1.28  $&$  1 $&    26.6 &$  0.56  $&$   3.46  $&$   11.244 $&$   0.33 $&$   1.00 $&   0.18  &$     27.6 $&$   11.26  $&$   0.34 $&$   1.04  $ \\   
06191224$+$2810044 &  184.303 &$   6.043$ &$    0.473$  &  1 &$-$& $-$ &  g  &$-$ &$-$ & *  &  D3 &  c  & o   & $- $ & $-$&  $-$ & $-$  &$   11.419  $&$  0.34  $&$  1.19  $&$  1 $&    28.8 &$  0.90  $&$   3.49  $&$   11.243 $&$   0.26 $&$   0.96 $&   0.15  &$     29.7 $&$   11.25  $&$   0.27 $&$   0.99  $ \\   
07003437$-$1020151 &  223.172 &$  -2.690$ &$    0.763$  &  1 &$-$& $-$ &  g  &$-$ & N  & *  &  D1 & $-$ &$- $ &  h   &  U &  $-$ & $-$  &$   11.533  $&$  0.40  $&$  1.37  $&$  1 $&    29.0 &$  0.80  $&$   3.76  $&$   11.249 $&$   0.28 $&$   0.99 $&   0.24  &$     30.9 $&$   11.25  $&$   0.29 $&$   1.04  $ \\   
07070637$-$2237435 &  234.885 &$  -6.853$ &$    0.271$  &  1 &$-$& $-$ &  g  &$-$ &$-$ & *  &  D1 &  c  & o   & $- $ & $-$&  $-$ & $-$  &$   11.347  $&$  0.34  $&$  1.12  $&$  1 $&    28.6 &$  0.62  $&$   3.67  $&$   11.246 $&$   0.30 $&$   0.99 $&   0.08  &$     29.0 $&$   11.25  $&$   0.30 $&$   1.01  $ \\   
07161385$-$2703250 &  239.807 &$  -6.998$ &$    0.416$  &  1 &$-$& $-$ &  g  &$-$ &$-$ & *  &  D4 &  c  & o   & $- $ & $-$&  $-$ & $-$  &$   11.400  $&$  0.38  $&$  1.27  $&$  1 $&    26.6 &$  0.84  $&$   3.68  $&$   11.245 $&$   0.31 $&$   1.06 $&   0.13  &$     27.3 $&$   11.26  $&$   0.32 $&$   1.09  $ \\   
07351136$-$2633207 &  241.353 &$  -3.027$ &$    0.790$  &  1 &$-$& $-$ &  g  & t  & N  & *  &  D1 & $-$ &$- $ & $- $ & $-$&  $-$ & $-$  &$   11.544  $&$  0.41  $&$  1.26  $&$  1 $&    37.0 &$  0.32  $&$   3.82  $&$   11.250 $&$   0.28 $&$   0.87 $&   0.24  &$     39.4 $&$   11.26  $&$   0.29 $&$   0.92  $ \\   
08115921$-$2433386 &  243.913 &$   5.114$ &$    0.173$  &  1 &$-$& $-$ &  g  &$-$ &$-$ & *  &  D4 &  c  & o   & $- $ &  V &  $-$ & $-$  &$   11.309  $&$  0.33  $&$  1.06  $&$  1 $&    30.0 &$  0.82  $&$   3.64  $&$   11.245 $&$   0.30 $&$   0.98 $&   0.05  &$     30.2 $&$   11.25  $&$   0.31 $&$   0.99  $ \\   
10005300$-$4241045 &  272.569 &$   9.912$ &$    0.226$  &  1 &$-$& $-$ &  g  & t  &$-$ & *  &  D1 &  c  & o   & $- $ &  V &  $-$ & $-$  &$   11.329  $&$  0.35  $&$  1.15  $&$  1 $&    50.0 &$  0.28  $&$   3.52  $&$   11.245 $&$   0.31 $&$   1.04 $&   0.07  &$     50.6 $&$   11.25  $&$   0.31 $&$   1.05  $ \\   
12133964$-$5618483 &  297.680 &$   6.173$ &$    0.471$  &  1 &$-$& $-$ &  g  &$-$ &$-$ & *  &  D1 &  c  & o   & $- $ &  V &  $-$ & $-$  &$   11.421  $&$  0.36  $&$  1.20  $&$  1 $&    25.4 &$  0.64  $&$   3.91  $&$   11.246 $&$   0.29 $&$   0.96 $&   0.15  &$     26.2 $&$   11.26  $&$   0.30 $&$   0.99  $ \\   
13015676$-$5751317 &  304.335 &$   4.984$ &$    0.511$  &  1 &$-$& $-$ &  g  &$-$ & P  & *  &  D1 & $-$ &$- $ & $- $ &  V &  $-$ & $-$  &$   11.431  $&$  0.49  $&$  1.40  $&$  1 $&    27.6 &$  0.54  $&$   4.05  $&$   11.241 $&$   0.41 $&$   1.15 $&   0.16  &$     28.5 $&$   11.25  $&$   0.42 $&$   1.18  $ \\   
15315422$-$4957230 &  327.527 &$   5.095$ &$    0.559$  &  2 &$-$& $-$ &  g  &$-$ &$-$ & *  &  D1 &  c  & o   & $- $ &  V &  $-$ & $-$  &$   11.450  $&$  0.35  $&$  1.26  $&$  1 $&    24.0 &$  0.74  $&$   4.21  $&$   11.242 $&$   0.26 $&$   0.98 $&   0.17  &$     24.9 $&$   11.25  $&$   0.27 $&$   1.02  $ \\   
15410922$-$6242284 &  321.054 &$  -5.970$ &$    0.414$  &  1 &$-$& $-$ &  g  &$-$ &$-$ & *  &  D4 &  c  & o   & $- $ & $-$&  $-$ & $-$  &$   11.397  $&$  0.30  $&$  1.11  $&$  1 $&    32.2 &$  0.80  $&$   4.13  $&$   11.243 $&$   0.23 $&$   0.91 $&   0.13  &$     33.1 $&$   11.25  $&$   0.24 $&$   0.93  $ \\   
16221431$-$3809491 &  342.120 &$   8.180$ &$    1.046$  &  1 &$-$& $-$ &  g  &$-$ & N  & *  &  D3 & $-$ &$- $ & $- $ &  V &  $-$ & $-$  &$   11.636  $&$  0.40  $&$  1.34  $&$  1 $&    19.2 &$  0.58  $&$   3.99  $&$   11.246 $&$   0.23 $&$   0.82 $&   0.32  &$     20.9 $&$   11.26  $&$   0.25 $&$   0.89  $ \\   
16475970$-$3330049 &  349.139 &$   7.412$ &$    0.726$  &  1 &$-$& $-$ &  g  &$-$ & P  & *  &  D1 & $-$ &$- $ & $- $ &  V &  $-$ & $-$  &$   11.517  $&$  0.35  $&$  1.46  $&$  1 $&    26.2 &$  0.74  $&$   4.19  $&$   11.246 $&$   0.23 $&$   1.10 $&   0.22  &$     27.8 $&$   11.25  $&$   0.25 $&$   1.15  $ \\   
18261102$+$0058105 &   30.910 &$   6.060$ &$    1.600$  &  1 &$-$& $-$ &  g  &$-$ & N  & *  &  D2 &  e  & o   & $- $ & $-$&  $-$ & $-$  &$   11.842  $&$  0.39  $&$  1.51  $&$  1 $&    22.6 &$  0.70  $&$   4.04  $&$   11.246 $&$   0.13 $&$   0.71 $&   0.50  &$     26.3 $&$   11.25  $&$   0.16 $&$   0.82  $ \\   
21014752$+$5743323 &   95.949 &$   7.445$ &$    0.849$  &  1 &$-$& $-$ &  g  &$-$ &$-$ & *  &  D1 &  c  & o   & $- $ & $-$&  $-$ & $-$  &$   11.564  $&$  0.46  $&$  1.42  $&$  1 $&    22.8 &$  0.74  $&$   3.77  $&$   11.248 $&$   0.32 $&$   1.00 $&   0.26  &$     24.4 $&$   11.26  $&$   0.34 $&$   1.05  $ \\   
21283297$+$3728032 &   84.364 &$  -9.777$ &$    0.290$  &  1 &$-$& $-$ &  g  &$-$ &$-$ & *  &  D3 &  c  & o   & $- $ & $-$&  $-$ & $-$  &$   11.350  $&$  0.37  $&$  1.11  $&$  1 $&    29.4 &$  0.76  $&$   3.53  $&$   11.242 $&$   0.32 $&$   0.96 $&   0.09  &$     29.9 $&$   11.25  $&$   0.33 $&$   0.98  $ \\   
22382102$+$5031104 &  102.359 &$  -6.959$ &$    0.225$  &  1 &$-$& $-$ &  g  &$-$ &$-$ & *  &  D1 &  c  & o   & $- $ & $-$&  $-$ & $-$  &$   11.328  $&$  0.51  $&$  1.27  $&$  1 $&    33.6 &$  0.72  $&$   3.66  $&$   11.244 $&$   0.47 $&$   1.15 $&   0.07  &$     34.0 $&$   11.25  $&$   0.48 $&$   1.17  $ \\   
\end{longtable}
\end{landscape}
\addtocounter{table}{-1}
\begin{landscape}
%\begin{longtable}{lrrrcclccccccllcclrrrrrrrrrrrrrrr}
\begin{longtable}{@{\extracolsep{1.0mm}}l@{\extracolsep{1.0mm}}r@{\extracolsep{1.0mm}}r@{\extracolsep{1.0mm}}r@{\extracolsep{1.0mm}}c@{\extracolsep{1.0mm}}c@{\extracolsep{1.0mm}}l@{\extracolsep{1.0mm}}c@{\extracolsep{1.0mm}}c@{\extracolsep{1.0mm}}c@{\extracolsep{1.0mm}}c@{\extracolsep{1.0mm}}c@{\extracolsep{1.0mm}}c@{\extracolsep{1.0mm}}l@{\extracolsep{1.0mm}}l@{\extracolsep{1.0mm}}c@{\extracolsep{1.0mm}}c@{\extracolsep{1.0mm}}l@{\extracolsep{1.0mm}}r@{\extracolsep{2.0mm}}r@{\extracolsep{2.0mm}}r@{\extracolsep{1.0mm}}r@{\extracolsep{1.0mm}}r@{\extracolsep{2.0mm}}r@{\extracolsep{2.0mm}}r@{\extracolsep{2.0mm}}r@{\extracolsep{1.0mm}}r@{\extracolsep{1.0mm}}r@{\extracolsep{1.0mm}}r@{\extracolsep{1.0mm}}r@{\extracolsep{1.0mm}}r@{\extracolsep{1.0mm}}r@{\extracolsep{1.0mm}}r}
\caption{{\normalsize b) 2MASS EBV exclusion sample } \label{ebvexcltab}}\\
\hline     
\noalign{\smallskip}
2MASX J          &  $l$ &  $b$ & EBV & \multicolumn{3}{c}{Object} & \multicolumn{4}{c}{Sample} & Disc  & 2MRS & \multicolumn{2}{c}{Velocity} & NIR  & Photom & Ext  & $K_{20}$ & $H$-$K$ & $J$-$K$ & vc  & $a$ & $b/a$ & st.d. & \ko & ($H$-$K)^o$ & ($J$-$K)^o$  &$\ak$& $a^{\rm d}$& \kd &($H$-$K)^{o,c}$ & ($J$-$K)^{o,c}$ \\ 
                 &  deg &  deg & mag & class & off & flg          &  gal & TF   & obs  & ext'd & type  &      & Opt    & HI                  & flg  & flg    & flg  & mag      & mag     &  mag    &     & ''  &       &       & mag & mag &  mag & mag & $''$       & mag & mag &  mag \\
(1)              & (2a) & (2b) & (3) & (4)   & (5) & (6)          & (7a) & (7b) & (7c) & (7d)  & (8)   & (9)  & (10a)  & (10b)               & (11) & (12)   & (13) & (14)     & (15)    & (16)    &(17) & (18)& (19)  & (20)  & (21)& (22)& (23) & (24)& (25)       & (26)& (27) & (28) \\
\noalign{\smallskip}
\hline     
\noalign{\smallskip}
\endfirsthead
\caption{\normalsize b) continued.}\\
\hline     
\noalign{\smallskip}
2MASX J          &  $l$ &  $b$ & EBV & \multicolumn{3}{c}{Object} & \multicolumn{4}{c}{Sample} & Disc  & 2MRS & \multicolumn{2}{c}{Velocity} & NIR  & Photom & Ext  & $K_{20}$ & $H$-$K$ & $J$-$K$ & vc  & $a$ & $b/a$ & st.d. & \ko & ($H$-$K)^o$ & ($J$-$K)^o$  &$\ak$& $a^{\rm d}$& \kd &($H$-$K)^{o,c}$ & ($J$-$K)^{o,c}$ \\ 
                 &  deg &  deg & mag & class & off & flg          &  gal & TF   & obs  & ext'd & type  &      & Opt    & HI                  & flg  & flg    & flg  & mag      & mag     &  mag    &     & ''  &       &       & mag & mag &  mag & mag & $''$       & mag & mag &  mag \\
(1)              & (2a) & (2b) & (3) & (4)   & (5) & (6)          & (7a) & (7b) & (7c) & (7d)  & (8)   & (9)  & (10a)  & (10b)               & (11) & (12)   & (13) & (14)     & (15)    & (16)    &(17) & (18)& (19)  & (20)  & (21)& (22)& (23) & (24)& (25)       & (26)& (27) & (28) \\
\noalign{\smallskip}
\hline     
\noalign{\smallskip}
\endhead   
\noalign{\smallskip}
\hline     
\endfoot   
%\noalign{\smallskip}
%ID                &    l     &$   b    $ &$   E(B-V)$  & xv & o &  ps & gal & tf &obs &ihb &  cl &tmrs & vo  & vh   &  U &   ph &  ext &$    K20    $&$  H-K   $&$  J-K   $&$ vc $&       a &$sup_ba  $&$stardens $&$   ko_as  $&$ hko_as $&$ jko_as $&  ak_sf  &$   a2d_sf $&$   kd_sf  $&$ hko_sf $&$  jko_sf $ \\   
03302716$+$3028296 &  159.115 &$ -21.004$ &$    1.445$  &  7 &$-$& $-$ & $-$ &$-$ &$-$ & *  & $-$ & $-$ &$- $ & $- $ &  U &  $-$ & $-$  &$   11.768  $&$  1.11  $&$  2.73  $&$  1 $&    15.6 &$  0.66  $&$   3.05  $&$   11.230 $&$   0.87 $&$   2.02 $&   0.45  &$   \cdots $&$  \cdots  $&$   0.90 $&$   2.11  $ \\  
\end{longtable}
\end{landscape}